\documentclass[apj]{emulateapj}

\newcommand{\ignore}[1]{}
\usepackage{apjfonts} \usepackage{amsmath,esint} \usepackage{lipsum}
\usepackage{ccaption}
\shorttitle{Vorticity-Preserving Scheme for Accretion Disks}
\shortauthors{Seligman \& Laughlin}

\begin{document} \title{A Vorticity-Preserving Hydrodynamical Scheme for Modeling Accretion Disk Flows}

\author{Darryl Seligman\altaffilmark{1}, Gregory Laughlin\altaffilmark{1}, }

\altaffiltext{1}{Department of Astronomy, Yale University, darryl.seligman@yale.edu} 

\begin{abstract}
 Vortices, turbulence, and unsteady non-laminar flows are likely both prominent and dynamically important features of astrophysical disks. Such strongly nonlinear phenomena are often difficult, however, to simulate accurately, and are generally amenable to analytic treatment only in idealized form. In this paper, we explore the evolution of compressible two-dimensional flows using an implicit dual-time hydrodynamical scheme that strictly conserves vorticity (if applied to simulate inviscid flows for which Kelvin's Circulation Theorem is applicable). The algorithm is based on the work of Lerat, Falissard \& Sid{\' e} (2007), who proposed it in the context of terrestrial applications such as the blade-vortex interactions generated by helicopter rotors. We present several tests of Lerat et al.'s vorticity-preserving approach, which we have implemented to second-order accuracy, providing side-by-side comparisons with other algorithms that are frequently used in protostellar disk simulations. The comparison codes include one based on explicit, second-order van-Leer advection, one based on spectral methods, and another that implements a higher-order Godunov solver. Our results suggest that Lerat et al's algorithm will be useful for simulations of astrophysical environments in which vortices play a dynamical role, and where strong shocks are not expected. 

\end{abstract}

\maketitle

\section{Introduction}\label{introduction}

Astrophysical models frequently appeal to unsteady nonlaminar flow in order to explain dynamical phenomena. Perhaps the best-known, and certainly the most-cited, example is the application of the \textit{alpha} prescription \citep{shakura1973} to model the viscous evolution of thin accretion disks.

Shakura and Sunayev's alpha prescription provides a simple parameterization of the disk's fluid viscosity, $\nu=\alpha c_s H$, in which $c_s$ is the sound speed, $H$ is the disk's scale height and the dimensionless constant $\alpha$ is posited to arise from turbulent processes. In the particular context of planet-forming disks, numerical values $\alpha \sim 0.001$ imply that a protostellar accretion disk spreads viscously on the observationally required  $\tau \sim R^2/\alpha c_s H \sim\,$Myr timescale, whereas the classical molecular viscosity implies extraordinarily long viscous timescales.

An extensive and sophisticated body of literature surrounds the nonlinear fluid processes that generate and maintain turbulence in protostellar disks, and which more generally facilitate long-term orbital angular momentum transfer. For recent reviews see, e.g. \citet{turner2014},  \citet{fromang2017}, and the references therein. Magnetorotational instabilities {e.g. }\citet{balbus1991}, self-gravitating spiral instabilities \citep{gammie2001},  Zombie Vortex Instabilities \citep{Marcus2015} and vortices driven by baroclinic or other hydrodynamical shear instabilities \citep{barge1995, adams1995, klahr2003} have all received extensive study.

At first glance, hydrodynamical turbulence generated by shear is a very attractive process to ascribe to protoplanetary disks. Turbulence (invariably accompanied by long-lived vortices) is observed in a variety of planetary atmospheres \citep{salby1984, marcus1988, bolton2017}, and indeed, unsteady flow is nearly ubiquitous in fluid contexts where both strong shear and large Reynolds numbers are present \citep{Abramowicz1992}. From a purely hydrodynamical standpoint, a compelling analogy supporting the alpha parameterization for effective viscosity in accretion disks is provided by the meteorological concept of \textit{eddy viscosity}. In Earth's atmosphere, complex nonlinear flow patterns such as von K\'{a}rm\'{a}n vortex streets -- that are known from laboratory experiments to manifest within only very narrow ranges of Reynolds number -- near $Re=VL/\nu\sim100$ -- are routinely observed \citep{vandyke1982}, where $V$ and $L$ are the characteristic fluid velocity and   length scale of the system. Large-scale atmospheric flows must therefore draw on an effective viscosity that is of order $10^{8}$ times larger than the molecular viscosity of air. Physically, the requisite diffusive transport is provided by coherent eddies (vortices) that act to transport momentum flux over macroscopic distances. It is natural to imagine that vortical disturbances play an analogous transport-enhancing role in protostellar disks.

\citet{kolmogorov1941} outlined a theory depicting fully developed three-dimensional fluid turbulence as an energetic cascade from large  vortices with dimension of order the system scale, $L$, down to eddies of size comparable to the mean free path, $\lambda_l$, of the individual molecules in the fluid. Adopting the assumption that the smallest scale turbulent structures are isotropic and invoking the conservation of kinetic energy per unit mass for each triad interaction of wavenumbers, Kolmogorov argued that at the onset of turbulence, the energy dissipation rate per unit mass $\epsilon$ of the largest vortices, with characteristic length scale $L$ and velocity $U$, is $\epsilon \sim {U^2}/({L/U}) \sim {U^3}/{L}$. In steady state, therefore, energy flowing into large vortices must be continously offloaded into smaller eddies with characteristic velocity $v_\lambda$ and length scale $\lambda$, such that
$\epsilon={v_\lambda^3}/{\lambda}$, implying a scaling relationship between the largest and smallest eddies, and defining the so-called \textit{turbulent cascade}, in which $v_\lambda=(\lambda/{L})^{1/3}U$.

When steady state has been established, the turbulent cascade extends from the system scale down to eddies of size $\lambda_l$ somewhat larger than the particle mean free path (generating the so-called inertial subrange). At scale $\lambda_l$, molecular viscosity converts coherent motions into heat. As a general rule, in a turbulent flow, when considering the energy distribution per unit log wavenumber $kE_k=dE/d\ln(k)$, the bulk of the perturbation energy is therefore contained in the largest scale vortices, while the vorticity is sequestered in the smallest eddies ($v_{\lambda}^2 \propto \lambda^{2/3}$, whereas $\Omega_{\lambda} \propto \lambda^{-2/3}$).

The aspect ratio, $h=c_{s}/v_{k} \ll 1$, where $v_{k}$ is the Keplerian velocity, of protostellar disks permits many of their important dynamical properties to be treated in the two-dimensional, or thin-disk approximation \citep{lyndenbell74}, and a somewhat imperfect analogy can be drawn with Earth's atmosphere. Over distances smaller than the $H\sim10\,\,{\rm km}$ scale height, the turbulent motions of tropospheric convection are fully three-dimensional, but at the planetary scale, turbulent flow is effectively constrained to two dimensions \citep{Charney71}. We note, however, that protostellar disk flow enforces supersonic shear over scales $\Delta R \gtrsim H$, a phenomenon not present in Earth's atmosphere.

Despite the obvious numerical advantage of working with one fewer dimensions, two-dimensional turbulence poses significant challenges to the connection of experimental results to analytic theory \citep[e.g.][]{vanheijst09}, and open questions remain, as strictly speaking, two-dimensional turbulence never manifests in nature or laboratory experiments. In two-dimensional incompressible flows, the mean-square vorticty of each fluid element is an inviscid constant of motion in addition to the kinetic energy per unit mass. Subsequently, the net transfer of any triad interaction of wavenumbers must be out of the middle wavenumber into \textit{both} the higher and lower wavenumbers, or vice versa \citep{Fjortoft53}. This is fundamentally different than three-dimensional turbulence, in which the vorticity tubes can warp to change the vorticity. In three dimensions, triad interactions can be treated like pair interactions where energy can flow in a one directional cascade. In fully developed two-dimensional turbulence, \citet{Kraichnan1967} demonstrated that the net transfer by each triad interaction must flow in both directions as a result of the vorticity constraint.\footnote{Interestingly, Kraichnan's findings stemmed from an investigation of Bose-Einstein condensates, in which the number density and kinetic-energy density of bosons play the same role as mean-square vorticity and kinetic energy.} He demonstrated that two-dimensional turbulence admits two independent energy spectra, $kE_k\sim \epsilon^{2/3}k^{-5/3}$ and $kE_k\sim \eta^{2/3}k^{-3}$ where $\epsilon$ is the flux of kinetic energy per unit mass across a given $k$, with units of energy per unit mass per unit time,  $\eta$ is the flux of mean-square vorticity across a given $k$, with units of vorticity squared per unit time, and $kE_k=d E/d\ln k$. He proved that the $-5/3$ range yields an energy cascade downward in wavenumber accompanied by zero vorticity cascade, while the $-3$ range yields a vorticity cascade upward in wavenumber accompanied by zero energy cascade. These two processes exist simultaneously, yielding the so-called dual cascade. 

Very high resolution direct numerical simulations \citep{boffetta2007, boffetta2010} provide a solid basis for the dual cascade scenario \citep{boffetta2012}, and the framework has found some support in laboratory experiments with thin soap films \citep[e.g.][]{bruneau2005}. In connection to astrophysical accretion disks, one might ask, if turbulence is responsible for the viscous dissipation, will it manifest globally as a two-dimensional dual cascade or only locally as a three-dimensional unidirectional cascade? In any event, either of these scalings (which both generate nonlinear structures of order the system size) are in \textit{fundamental tension} with any scenario in which an alpha viscosity acting at the grid or sub-grid scale coexists with global laminar flow. That is, turbulence implies unsteady features at every length scale, including the largest length scales. It is important to note that a flow with an alpha viscosity that arises from turbulent angular momentum transport is very different than a viscous laminar flow. A survey of the extant literature suggests that the potential applicability of the dual cascade framework to the context of thin astrophysical disks that exhibit near-Keplerian shear has not yet been thoroughly investigated.

While purely hydrodynamical global simulations of accretion disks do not tend to produce flows that are replete with vortices, there is a large literature that describes the various roles that disk vortices might play in the planet formation process, as evidenced by hundreds of citations to Barge \& Sommeria's (1995) article. A Keplerian disk, with its $\Omega(r)\sim r^{-3/2}$ angular velocity law has a specific angular momentum profile that increases with radius, $\partial_r(\Omega^2 r^2)>0$, and is thus linearly stable to perturbations via Rayleigh's well known criterion \citep{rayleigh1917}. It remains unclear whether finite-amplitude perturbations applied to flow with extremely high Reynolds number ($Re\gtrsim10^{10}$) associated with purely molecular viscosity will lead to sustained turbulence. Laboratory experiments involving quasi-Keplerian fluid columns wedged between rotating cylinders have given contradictory results. \citet{schartman2012} report laminar flow at Reynolds number $Re\sim2\times10^6$, whereas \citet{paoletti2012} find turbulent flow at $Re > 10^5$. Recent numerical work by \citet{shi2017} finds no evidence of sustained turbulence in models of these quasi-Keplerian flows at shear Reynolds numbers of order $Re \sim 10^5$.

Our work here fits into the larger effort to elucidate how the overarching inconsistency between laminar disk flow at large scales co-existing with two or three-dimensional turbulent flow on small scales is most properly resolved. For a fluid with $\vec{\nabla} P \times \vec{\nabla} \rho=0$ (i.e. for inviscid barotropic flow), Kelvin's circulation theorem \citep{landau1959} demands that the circulation $\Gamma$ around a closed surface, $S$, defined by
\begin{equation}\label{eq:circthm1}
\Gamma(t)=\oiint\limits_{S(V)} {\boldsymbol{\omega} }\cdot{\bf \hat{n}}\, d S\, ,
\end{equation}
where $\boldsymbol{\omega}=\vec{\nabla} \times \bf{u}$ is the vorticity, is conserved as the fluid is advected and otherwise deformed by the fluid flow,

\begin{equation}\label{eq:circthm2}
\frac{D}{Dt}\Gamma(t)=0\, .
\end{equation}

In protostellar disk simulations designed so that Equation \ref{eq:circthm2} should hold, vorticity is generally \textit{not} advectively conserved. Many numerical methods are formulated to evolve the continuity, momentum, and energy equations while preserving frontal transport of sharp large-scale features in the flow. The extent to which Kelvin's circulation theorem is respected, however, is rarely monitored. Numerical dissipation in Lagrangian schemes such as smooth particle hydrodynamics \citep{schafer2004} or in grid-based schemes such as FARGO \citep{masset2000, benitez2016} that have historically been used for simulations of protostellar disk processes such as planet migration \citep[see the comparisons of ][]{delvalborro2006}, strongly damp out grid-scale turbulence, thereby artificially enforcing laminar flows that are potentially at odds with the Kolmogorov and Kraichnan scalings. Higher-order methods for evolving the fluid equations, for example, those using Godunov-type solvers, as implemented in community codes such as FLASH \citep{fryxell2000}, PLUTO \citep{mignone2007},  PENCIL \citep{brandenburg2002}, or Athena++ \citep{white2016}, generally implement slope limiters or equivalent devices that maintain numerical stability, while simultaneously introducing a degree of numerical diffusion. 

In the computer graphics community, numerical simulations of nearly incompressible, moderate-to-high Reynolds number flows have constituted an area of intense study \citep[e.g.][]{elcott2007}, albeit with both a merit function and a motive that are quite different from those generally adopted by astronomers. In the context of an animation or a video game, a simulated fluid, be it smoke or a fiery explosion, needs to ``look'' right, and it often needs to update in real time. This requires that the simulated fluid display proper eddying behavior over a substantial range of length scales, and it requires that numerical dissipation  be kept to a low level (or actively corrected). 

A paper describing very simple, yet nonetheless innovative and influential semi-Lagrangian method, \textit{Stable Fluids} \citep{stam2001}, has received almost two thousand Google citations, while having gone simultaneously unmentioned in the astronomical literature. Stam's algorithm efficiently evolves visually realistic flows in $\vec{\nabla} \cdot {\bf v}=0$ fluids by removing numerically-induced compression, thereby exposing eddies with sizes ranging down to the grid scale. This task employs Helmholtz-Hodge decomposition, by which any sufficiently smooth vector field, ${\bf w}$, can be decomposed into the form

\begin{equation}\label{eq:HH}
    {\bf w}={\bf u} + \vec{\nabla} q\, ,
\end{equation}
where $\vec{\nabla} \cdot {\bf u}=0$ and $q$ is a scalar field. This decomposition separates a field, for example, a velocity field in a numerical simulation that has just received an advective update, $\partial {\bf u}/\partial t =({\bf u}\cdot \vec{\nabla}){\bf u}$, computed via finite differences, into an irrotational component and a purely circulating component. In practice, this is done by defining a projection operator ${\bf P}$,
\begin{equation}\label{eq:Project}
 {\bf P~w} = {\bf w} -\vec{\nabla} q={\bf u} \, ,
\end{equation}
that extracts the velocity field, ${\bf w}$'s divergence-free component, {\bf u}. The operator ${\bf P}$ is defined implicitly by the divergence of equation \ref{eq:Project},

\begin{equation}\label{eq:laplace}
\vec{\nabla}^2 q = \vec{\nabla} \cdot{\bf w} \, ,
\end{equation}
which constitutes a Poisson equation for $q$, solvable via standard techniques for elliptic PDEs. In practice, the projection of the advected velocity field onto its divergence-free component, accomplished by evaluating ${\bf u}={\bf w} - \vec{\nabla} q$, enforces incompressibility, while simultaneously providing an offset to the diffusive action of numerical dissipation at the grid scale. 

\begin{figure}
\begin{center}
\resizebox{0.25\textwidth}{!}{\includegraphics*[trim={8.5cm 1.9cm 8.5cm 1.9cm},clip]{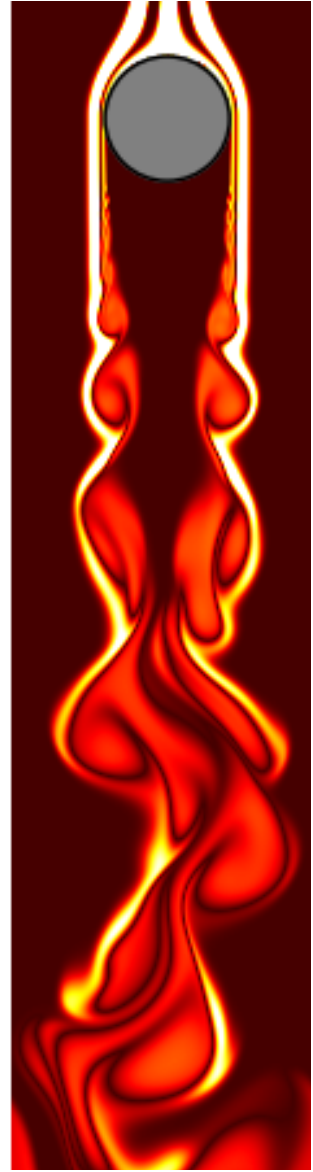}}
\end{center}
\caption{Flow past a cylinder computed with an implementation of Stam's \textit{Stable Fluids} algorithm after  $\sim 3$ vertical fluid  velocity crossing times on a 500x2000 zone grid. Color corresponds to the density (or concentration, given that $\vec{\nabla}\cdot \bf{v}=0$) of passively advected dye that is injected into the flow upstream of the cylinder. Despite using  first-order differencing, the implementation is able to produce visually realistic results by diverting fluid compression into eddying motions.}
\label{fig:Stable Fluids}
\end{figure}

Although the \textit{Stable Fluids} algorithm does not explicitly conserve either energy or vorticity, the application of the Helmholtz-Hodge decomposition does enforce mass conservation, and remarkably, it generates visually realistic, qualitatively accurate swirling flows (see Figure \ref{fig:Stable Fluids}).  It is also unconditionally stable in the sense that its semi-Lagrangian nature circumvents the Courant limit, so it is a very attractive procedure for quickly rendering the three-dimensional evolution of incompressible fluids.

It is surprising that an extremely simple first-order finite difference algorithm for incompressible fluids can yield qualitatively realistic results when the vorticity at the grid scale is \textit{constantly} modified as a consequence of applying a Helmholtz-Hodge decomposition to correct an imperfectly advected velocity field. The concurrent computational efficiency and high quality of these results has motivated our investigation of a very analogous, but more rigorous, approach for evolving mildly compressible fluids. In order to simulate high Reynolds number flows that do not suppress potentially critical non-laminar features, we have implemented a vorticity-preserving scheme for evolving Euler's formulation of the momentum equation, one that constantly enforces modification of the velocity field at the grid scale, much as Stable Fluids does.

The plan of this paper is as follows.  Having started with an extended outline of our motivation, we provide, in \S 2, a description of \citet{Lerat2007}'s vorticity-preserving algorithm (which they refer to as ``RBV2'', a nomenclature that we henceforth adopt). The treatment in Lerat et al.'s published description of RBV2 is slightly telegraphic. At the risk of being overly literal, we describe our implementation of the RBV2 scheme at a level of detail that should render it very straightforward for an interested reader to quickly reproduce. In \S 3, we outline three expository tests of the scheme. The first demonstrates that RBV2 can exactly and indefinitely preserve the propagation of a vortex with an analytically determined profile that is in exact geostrophic balance. The second test illustrates the evolution of a random, undriven field of subsonic velocity perturbations. The RBV2 algorithm's preservation of vorticity generates far less numerical diffusion of small-scale eddies, in strong contrast to the behaviour of the comparison codes. The third test demonstrates how RBV2 handles a shock tube. In \S 4, we demonstrate how RBV2 evolves vortices seeded into shearing sheet patches that approximate the base flow of near-Keplerian protostellar disks. A test of vortex merger suggests that RBV2 runs at a higher effective Reynolds number than the comparison schemes. In \S 5, we outline our optimism for the prospects of further use of RBV2 and conclude.


\section{A prescription for preserving vorticity in compressible flows}

In turbulent astrophysical flows, one expects that both significant compressions and persistent eddies will be present. The continuity equation
\begin{equation}\label{eq:continuity}
\frac{\partial{\rho}}{\partial t}=-\vec{\nabla}\cdot (\rho {\bf u})\, ,
\end{equation}
will thus dictate the evolution of the density in conjunction with the momentum equation 
\begin{equation}\label{eq:momentum}
\frac{\partial{\bf u}}{\partial t}+({\bf u} \cdot \vec{\nabla}){\bf u}=-\frac{\vec{\nabla} p}{\rho}-\vec{\nabla} \Phi +\nu\vec{\nabla}^2{\bf u}\, ,
\end{equation}
that controls the development of the flow's velocity. Assuming that the conventional molecular viscosity, $\nu$ is negligible, the curl of the momentum equation (equation \ref{eq:momentum})
gives an expression for the evolution of vorticity (${\bf \omega}=\vec{\nabla} \times {\bf u}$),
\begin{equation}\label{eq:vort}
\frac{\partial \boldsymbol{\omega}}{\partial t}=-({\bf u} \cdot \vec{\nabla})\boldsymbol{\omega}+(\boldsymbol{\omega}\cdot\vec{\nabla}) {\bf u} -(\vec{\nabla}\cdot {\bf u})\boldsymbol{\omega} + \frac{\vec{\nabla} \rho\times\vec{\nabla} p}{\rho^2}\, .
\end{equation}
which reduces to Eqn. \ref{eq:circthm2} for any barotropic, $P=P(\rho)$, equation of state.

In three dimensions, the vorticity in a Lagrangian fluid parcel can be modified by stretching and warping of vorticity tubes, compression and baroclinic effects stemming from $\vec{\nabla} \rho\times\vec{\nabla} p\neq0$. A numerical method must respect this equation in order to properly advect vorticity. This is no trivial task since at the grid scale, at the close of one timestep of evolution, the updated vorticity in a cell depends on the updated vorticities of each adjacent cell, necessitating an implicit approach.

In the mechanical engineering community, problems such as the blade-vortex interactions of helicopter rotors require numerical simulations that accurately track the evolution of vorticity, and which keep numerical diffusion of this quantity to a minimum. One approach to this problem consists of increasing the order (and hence the accuracy) of the discretization scheme, as outlined by \citet{abgrall2003} or \citet{bogey2004}. An alternate strategy has been set forth by \citet{morton2001}, who modeled sound propagation through a still medium. For purely acoustic fluid disturbances, in which Eqn. 2 is strictly true, \citet{morton2001} noted that the Lax-Wendroff scheme outlined by \citet{ni1982}, exactly conserves a discrete representation of the vorticity, and they modified Ni's approach to take specific advantage of its vorticity-preserving property. \citet{steinhoff1994} describe a ``vorticity confinement'' method in which a local term is added to the Euler equation that enforces vorticity conservation through the dynamic introduction of Lagrangian-like confined vortical regions. 

\citet{Lerat2007} further extended these ideas in the context of compressible flow. Their resulting
\textit{RBV2} algorithm despite being only second-order accurate does extremely well in test problems involving slowly advected vortices. From the standpoint of potential application to astrophysical problems, it is intriguing to note that it exactly preserves the hydrodynamical structure of a large-scale vortex in geostrophic balance that was seeded from an analytic solution. Such vortices (or close relatives thereof) are frequently invoked for processes such as dust-trapping in protostellar disks \citep[e.g.][]{surville2016}.

Adopting the notation of \citet{Lerat2007}, and specializing to a two-dimensional Cartesian geometry, the fluid continuity, momentum and energy equations written in vector form are
\begin{equation}\label{eq:vectorfluids}
    \frac{\partial \bf{w}}{\partial t}+ \frac{\partial \bf{f}}{\partial x} +\frac{\partial \bf{g}}{\partial y} =0\, ,
\end{equation}

for a fluid velocity defined by ${\bf u}=(u,v)$,

\begin{equation}\label{eq:w}
    \bf{w} = \begin{bmatrix}
           \rho \\
           \rho u \\
           \rho v \\
           \rho E
         \end{bmatrix}\, ,
\end{equation}
\begin{equation}\label{eq:f}
    \bf{f}=\bf{f}(w) = \begin{bmatrix}
           \rho u \\
           \rho u^2 + p \\
           \rho uv \\
           (\rho E + p)u
         \end{bmatrix}\, ,
\end{equation}
and
\begin{equation}\label{eq:g}
    \bf{g}=\bf{g}(w) = \begin{bmatrix}
           \rho v \\
           \rho uv \\
           \rho v^2 +p \\
           (\rho E +p)v
         \end{bmatrix}\, .
\end{equation}

Time-stepped finite difference representations of Equations \ref{eq:vectorfluids}-\ref{eq:g} experience numerical dissipation, either that which is explicitly added, or that which is induced by the upwinding used to compute advection. As a consequence, the representation of fluid motion is compromised in transitioning from a continuous description to a discrete, grid-based regime.
To clarify, consider a continuous model PDE with explicit dissipation,

\begin{equation}\label{eq:continuousscheme}
\frac{\partial {\bf w}}{\partial t} +\frac{\partial  {\bf f}}{\partial x} +\frac{\partial {\bf g}}{\partial y} =\epsilon(\frac{\partial^2 {\bf w}}{\partial x^2}+\frac{\partial^2 {\bf w}}{\partial y^2})\, ,
\end{equation}
where $\epsilon$ is some positive parameter. In such a situation, the diffusive evolution of vorticity is given by
\begin{equation}\label{eq:vortlaplace}
\frac{\partial {\boldsymbol\omega}}{\partial t}  = \epsilon \vec{\nabla}^{2} {\boldsymbol \omega}\, .
\end{equation}
The ingenuity of Lerat et al.'s method stems from constructing dissipative terms, $\epsilon \vec{\nabla}^{2} {\boldsymbol \omega}$, that exactly respect the vorticity equation  -- that is, equal the right hand side of Equation 8. 

A finite difference representation of the following PDE,

\begin{equation}\label{eq:dissipativescheme}
\frac{\partial {\bf w}}{\partial t} +\frac{\partial {\bf f}}{\partial x} +\frac{\partial {\bf g}}{\partial y}=\epsilon_1\frac{\partial(\Phi_1{\bf q})}{\partial x} +\epsilon_2\frac{\partial(\Phi_2 {\bf q})}{\partial y}\, ,
\end{equation}
is proven to be vorticity preserving in Theorem 1 of  \citet{Lerat2007}, when $\epsilon_1$, $\epsilon_2$ are positive parameters, ${\bf q}$ is a vector involving odd derivatives in space and time of $\bf{w}$, $\bf{f}$, and $\bf{g}$,  and  $\Phi_1, \Phi_2$ are constant matrices with the same eigenvectors as the Jacobian matrices $A = d{\bf f}/dw$ and $B = d {\bf g}/dw$ respectively.

To implement the vorticity-preserving finite scheme that approximates Eqn. 15 on a Cartesian mesh with $x_j=j\delta x$,  $y_k=k\delta y$,  and time steps $t^n=n\Delta t$, we first define the one dimensional difference and average operators, for an arbitrary primitive variable $\psi$,

\begin{equation}\label{eq:d1}
(\delta_1 \psi)_{j+\frac12,k} = \psi_{j+1,k}-\psi_{j,k}\, ,
\end{equation}
\begin{equation}\label{eq:d2}
(\delta_2 \psi)_{j,k+\frac12} = \psi_{j,k+1}-\psi_{j,k}\, ,
\end{equation}

\begin{equation}\label{eq:mu1}
(\mu_1 \psi)_{j+\frac12,k}=\frac12(\psi_{j+1,k}+\psi_{j,k})\, ,
\end{equation}
and

\begin{equation}\label{eq:mu2}
(\mu_2 \psi)_{j,k+\frac12}=\frac12(\psi_{j,k+1}+\psi_{j,k})\, ,
\end{equation}
and the time difference operators,

\begin{equation}\label{eq:Dv}
(\Delta \psi)^n=\psi^{n+1}-\psi^n\, ,
\end{equation}
and
\begin{equation}\label{eq:Dvn1}
(\bar{\Delta} \psi)^{n+1}=\frac32 \psi^{n+1} - 2\psi^{n} +\frac12 \psi^{n-1}\, .
\end{equation}

With these operators, the second-order (in space and time) residual based finite-difference scheme that Lerat and Corre refer to as RBV2, is defined by
\begin{equation}\label{eq:RBV2}
    (\Lambda {\bf\widetilde{r}}) ^{n+1}_{j,k} = 0 \, ,
\end{equation}
and has been proven to be vorticity preserving for the full Euler equations (Theorem 7 of \citet{Lerat2007}). In the above equation, the discrete residual ${\bf\widetilde{r}}$ is defined as
\begin{equation}\label{eq:rtw}
    {\bf\widetilde{r}}^{n+1}_{j+\frac12,k+\frac12}=(\mu_1\mu_2 \frac{\bar{\Delta} }{\Delta t} {\bf w}+\frac{\delta_1\mu_2}{\delta x} {\bf f} +\frac{\delta_2 \mu_1}{\delta y} {\bf g})^{n+1}_{j+\frac12,k+\frac12}\, ,
\end{equation}
and the difference operator $\Lambda$  is given by

\begin{equation}\label{eq:lambda}
\Lambda = \mu_1\mu_2 -\frac12 (\delta_1\mu_2\Phi_1+\delta_2\mu_1\Phi_2)\, .
\end{equation}
For a closure relation connecting pressure and density, a polytropic equation of state
  \begin{equation}\label{eq:eos}
  p=K\rho^\gamma\, ,
  \end{equation}
is assumed.

As formulated, the method is fully implicit, and can be cast in the form 
 
  \begin{equation}\label{eq:RBV2solve}
  \begin{split}
  \frac{ \bar{\Delta}}{\Delta t} \mu_1\mu_2\mu_1\mu_2 {\bf w}+ \frac{1}{\delta x} \delta_1 \mu_2 \mu_1\mu_2{\bf f}+ \frac{1}{\delta y} \delta_2 \mu_1  \mu_1\mu_2 {\bf g}=\\\frac{\delta x}{2} \frac{1}{\delta x} \delta_1 \mu_2 (A {\bf\widetilde{r}})+\frac{\delta y}{2} \frac{1}{\delta y} \delta_2 \mu_1(B {\bf\widetilde{r}})\, ,
  \end{split}
  \end{equation}
 Definition 2 \citep{Lerat2007}. The left hand side of the above expression involving ${\bf f}$ and ${\bf g}$ corresponds to the convective residuals and the right hand side corresponds to the dissipative residuals. As a consequence of the implicit formulation, the solution vector at the forward time step, ${\bf w}^{n+1}$, is itself a function of the solution. Therefore, approximate solutions such as alternate-line relaxation of the Jacobi or Gauss-Seidel relaxation can be used.  
 
 In our implementation, we solve the residual equation  using a dual time stepping technique \citep{Jameson1991}. The method involves deriving a finite volume scheme by applying the relevant equations to control volumes to get a set of  ordinary differential equations of the form,

\begin{equation}\label{eq:ode}
    \frac{d}{dt}({\bf w} V)+{\bf R}({\bf w})=0 \, ,
\end{equation}
where $V$ is the cell volume and ${\bf R}({\bf w})$ is the discrete residual. This equation can be treated as a modified steady state problem and solved with fictitious time steps $t^*$,
\begin{equation}\label{eq:t*}
    \frac{\partial{\bf w}}{\partial t^*} +{\bf R}^*({\bf w})=0 \, .
\end{equation}

The residual is split into its convective and dissipative terms,

\begin{equation}\label{eq:res}
{\bf R}^*({\bf w})={\bf Q}({\bf w})+{\bf D}({\bf w})\, .
\end{equation}
In the context of RBV2, these residuals are defined as

\begin{equation}\label{eq:Q}
    {\bf Q}({\bf w})=\frac{1}{\delta x} \delta_1 \mu_2 \mu_1\mu_2{\bf f}+ \frac{1}{\delta y} \delta_2 \mu_1  \mu_1\mu_2 {\bf g}\, ,
\end{equation}
and
\begin{equation}\label{eq:D}
    {\bf D}({\bf w})=-[\frac{\delta x}{2} \frac{1}{\delta x} \delta_1 \mu_2 (A {\bf\widetilde{r}})+\frac{\delta y}{2} \frac{1}{\delta y} \delta_2 \mu_1(B {\bf\widetilde{r}})]\, .
\end{equation}
 The solution vector at subsequent time steps ${\bf w}^{n+1}$ is found by iterating over the following algorithm, where the superscript k symbolizes successive guess for ${\bf w}$,

\begin{equation}\label{eq:guess1}
{\bf w}^{n+1,0}={\bf w}^n \, ,
\end{equation}

and successive guesses are

\begin{equation}\label{eq:wguess}
{\bf w}^{n+1,k}={\bf w}^n-\alpha_k\Delta t^*[{\bf Q}^{k-1}+{\bf D}^{k-1}]\, .
\end{equation}

The initial convective and dissipative residuals are calculated from the previous time step, 
\begin{equation}\label{eq:Q0}
{\bf Q}^0={\bf Q}[{\bf w}^n]\, ,
\end{equation}
and
\begin{equation}\label{eq:D0}
{\bf D}^0={\bf D}[{\bf w}^n]\, .
\end{equation}
The kth residuals are calculated using
\begin{equation}\label{eq:Qk}
{\bf Q}^k={\bf Q}[{\bf w}^{n+1,k}]\, ,
\end{equation}
and

\begin{equation}\label{eq:Dk}
{\bf D}^k=\beta_k{\bf D}[{\bf w}^{n+1,k}]+(1-\beta_k){\bf D}^{k-1} \, .
\end{equation}

The coefficients $\alpha_k$ and $\beta_k$ are chosen to maximize the stability interval along the imaginary and negative real axis of each Fourier mode of the solution. \citet{Jameson1991} reported  that  effective choice for $\alpha$ and $\beta$ are $\alpha_1=1/3$, $\alpha_2=4/15$,  $\alpha_3=5/9$, $\alpha_4=1$ and $\beta_1=1$, $\beta_2=1/2$, $\beta_3=0$, and $\beta_4=0$.
\begin{figure*}
\begin{center}
\resizebox{1.09\textwidth}{!}{\includegraphics*{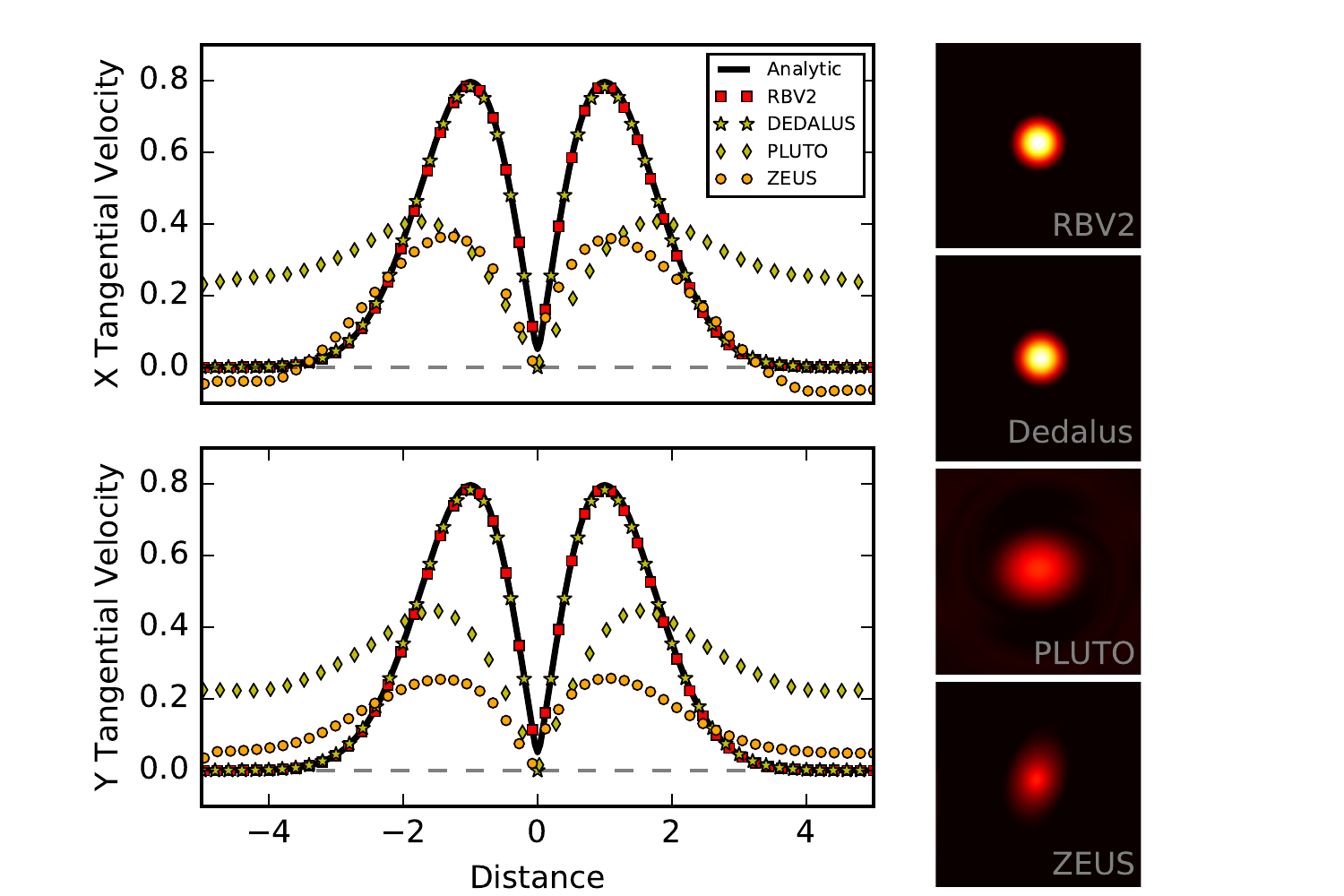}}
\end{center}
\caption{\textit{Left Panels}: \textit{x}-directional and \textit{y}-directional cross sections of the tangential velocity of the seeded Yee et al. (1999) vortex following significant hydrodynamical evolution. The seeded, analytic velocity profile is shown in the black solid line, and is unchanging in time. The red squares, yellow stars, yellow diamonds and orange circles show the vortex's velocity profile after 100 orbits as evolved with the RBV2, Dedalus, PLUTO (WENO+RK3) and ZEUS (van Leer) codes respectively. \textit{Right Panels}: color-coded maps of the vorticity after 100 orbits of the entire vortex in RBV2 (upper right), Dedalus (upper middle right) PLUTO (lower middle right) and ZEUS (lower right). The initial vorticity is almost indistinguishable from that shown in RBV2 and Dedalus, and the maximum vorticity in the initial and RBV2, Dedalus, PLUTO and ZEUS evolved vortex is 2.62, 2.62, 2.62, 1.05 and 1.09 respectively. All simulations were run with 256x256 zones and PLUTO was run with RK3 time stepping and the WENO3 (weighted essentially non-oscillatory) reconstruction scheme \citep{liu1994}. Dedalus was run solving the Navier-Stokes equation with a kinematic viscosity $\nu\sim1x10^{-5}$. While Dedalus and RBV2 both maintain the analytic solution to high fidelity, RBV2 is computationally faster by a factor of ten.}
\label{fig:Analytic Vortex}
\end{figure*}

As noted by \citep{Fallisard2008}, in the case of diagonal flows, spurious short wavelength perturbations may manifest in the crosswise direction, and we observed such features in an unmodified implementation of the scheme outlined by Lerat et al. (2007). In order to dissipate these perturbations while maintaining the desirable vorticity-preserving quality of RBV2, high order spatial filters \citep{lele1992} are applied at the end of each simulation time step. 

The one-dimensional filter as described in Appendix C.2 of \citet{lele1992} for a strictly periodic domain is 
\begin{equation}\label{eq:LeleFilter}
\begin{split}
    \beta \hat{f}_{i-2}+\alpha\hat{f}_{i-1}+\hat{f}_{i}+\alpha \hat{f}_{i+1}+\beta\hat{f}_{i+2}=a f_i +\frac{b}{2}(f_{i+1}+f_{i-1}) \\\  + \frac{c}{2}(f_{i+2}+f_{i-2})+\frac{d}{2}(f_{i+3}+f_{i-3})\, ,
\end{split}
\end{equation}
where $\hat{f}_i$/$f_i$ denote  the filtered/unfiltered value at mesh point $i$. As noted in \citet{lele1992}, $\alpha = 0.6522474$, $\beta=0.1702929$, $a=0.9891856$, $b=1.321180$, $c=0.3333548$, and $d=0.001359850$. This filter is straightforward to cast in the form of a matrix inversion problem. We implement such a filter for both the $x$-directional and $y$-directional flows at each timestep, and alternate the order of application every timestep to prevent asymmetries. 

With non-perioidic boundary conditions, near the boundaries the filter takes the form

\begin{equation}\label{eq:LeleBoundary1}
\hat{f}_{1}=\frac{15}{16}f_1 +\frac{4}{16}f_2-\frac{6}{16}f_3+\frac{4}{16}f_4-\frac{1}{16}f_5\, ,
\end{equation}

\begin{equation}\label{eq:LeleBoundary1}
\hat{f}_{2}=\frac{3}{4}f_2 +\frac{1}{16}f_1+\frac{6}{16}f_3-\frac{4}{16}f_4+\frac{1}{16}f_5\, ,
\end{equation}

\begin{equation}\label{eq:LeleBoundary1}
\hat{f}_{3}=\frac{5}{8}f_3 -\frac{1}{16}f_1+\frac{4}{16}f_2+\frac{4}{16}f_4-\frac{1}{16}f_5\, ,
\end{equation}

where $\hat{f}_{1}$ and $\hat{f}_{2}$ represent the first and second ghost zones, and $\hat{f}_{3}$ is the first zone in the computational domain.

\section{Hydrodynamical Tests}

Having described our implementation of the RBV2 method, we move to consider a number of tests. Because the method is designed to exactly conserve vorticity, we expect that it will register its most impressive performance in connection with flows that are dominated by vortical motion. Conversely, we expect irrotational flows, particularly those presenting shocks or sharp fronts, to be evolved with less aplomb, especially in comparison to conventional high-order or Godunov schemes. 

\subsection{Hydrodynamical Vortex Dissipation}

We first track the fidelity with which the RVB2 scheme evolves an analytic vortex solution to the inviscid fluid equations. Similar vortex tests were presented by \citet{Springel2010} rendered with the codes AREPO and ATHENA  \citep{Stone2008}. We use the solution proposed by Yee et al. (1999), in which a geostrophically balanced vortex, propagating in an ideal gas, is defined by a radial temperature profile,

\begin{equation}\label{eq:temp}
    T=1-\frac{(\gamma-1)\Gamma^2}{8\gamma\pi^2}\exp(1-r^2)\, ,
\end{equation}
and  velocity field ${\bf U}=(u,v)$, with
\begin{equation}\label{eq:vx}
u=-\frac{\Gamma}{2\pi}y\exp(\frac{1-r^2}{2})\, ,
\end{equation}
and
\begin{equation}\label{eq:vy}
v=\frac{\Gamma}{2\pi}x\exp(\frac{1-r^2}{2})\, ,
\end{equation}
superimposed on an undisturbed constant density background. In the above, $r^2=x^2+y^2$, the vortex strength parameter is set to $\Gamma=5$ and the ratio of specific heats is $\gamma=1.4$. Note that $\Gamma$ and $\gamma$ have now been redefined and have different meanings than have been used earlier in the paper. The ideal gas law gives the thermodynamic equation of state,
\begin{equation}\label{eq:eos2}
p=\rho T\, ,
\end{equation}
and by virtue of uniform entropy throughout the vortex,
\begin{equation}\label{eq:rhoT}
\rho=T^{\frac{1}{\gamma-1}}\, .
\end{equation}
With steady state arising from geostrophic balance, the pressure gradient everywhere balances the centripetal acceleration,

\begin{equation}\label{eq:balance}
|\frac{\vec{\nabla} P}{\rho}| = \frac{|{\bf U}^2|}{r}\, ,
\end{equation}
so that component-wise
\begin{equation}\label{eq:Px}
\frac{\partial P}{\partial x} = \frac{\Gamma^2}{4\pi^2}\exp(1-r^2) x \rho\, ,
\end{equation}

\begin{equation}\label{eq:Py}
\frac{\partial P}{\partial y} = \frac{\Gamma^2}{4\pi^2}\exp(1-r^2) y \rho\, ,
\end{equation}
and 

\begin{equation}\label{eq:V2}
|{\bf U}|^2 = \frac{\Gamma^2}{4\pi^2}r^2\exp(1-r^2) \, .
\end{equation}

Given that RBV2 is formally second-order accurate in space, it is useful to compare its performance with that of a frequently used second-order conventional upwinding hydrodynamical scheme. We thus seed the Yee et al. vortex solution into  a purpose-written hydrodynamics code that uses a van Leer second order upwind advection scheme \citep{vanleer1977}, as implemented in ZEUS \citep{stone1992} and FARGO \citep{masset2000} and for the radial and z-direction momentum advection in the more recent FARGO3D code \citep{benitez2016}. We also simulate the evolution of the flow with the PLUTO code \citep{mignone2012}, run with third-order Runge-Kutta time stepping and the WENO3 reconstruction scheme and with Dedalus \citep{Burns2016}, an open source spectral code for solving differential equations. Specifically, we implemented a solution of the Navier-Stokes equation with a kinematic viscosity $\nu\sim1x10^{-5}$, corresponding to a highly inviscid flow. 

For comparison purposes, RBV2, Dedalus, ZEUS and PLUTO are all run at identical 256x256 grid resolution. Based on numerical experiment, we determined that in order to simulate high Reynolds number flow with the spectral method,  Dedalus required a timestep that was one tenth of that used in the other three codes.  As presented, the Dedalus simulation took 10 times longer to run than RBV2.

Figure \ref{fig:Analytic Vortex}  compares the analytically described tangential velocity profile of the vortex with the profiles exhibited by the four methods after they have evolved the flow for a time corresponding to 100 vortex orbits, where one vortex orbit is defined by the amount of simulation time it takes for a fluid parcel in the vortex to circulate once at the radius of maximum tangential velocity. It is evident from the figure that ZEUS and PLUTO do not respect Kelvin's circulation theorem, whereas RBV2 and Dedalus perfectly retain the original vortex structure to a startling degree. Although this test case is certainly idealized, and certainly favors a vorticity-preserving method by stipulating a flow dominated by a static vortex, it is evident that RBV2 and Dedalus merit strong consideration for modeling flows in which both rotational and irrotational motion are present, although RBV2 is capable of running at higher computational efficiency than Dedalus. 

\subsection{Evolution of Randomly Perturbed Subsonic Flow}
\begin{figure*}
\begin{center}
\resizebox{0.22\textwidth}{!}{\includegraphics*[trim={2.5cm .1cm 2.5cm .1cm},clip]{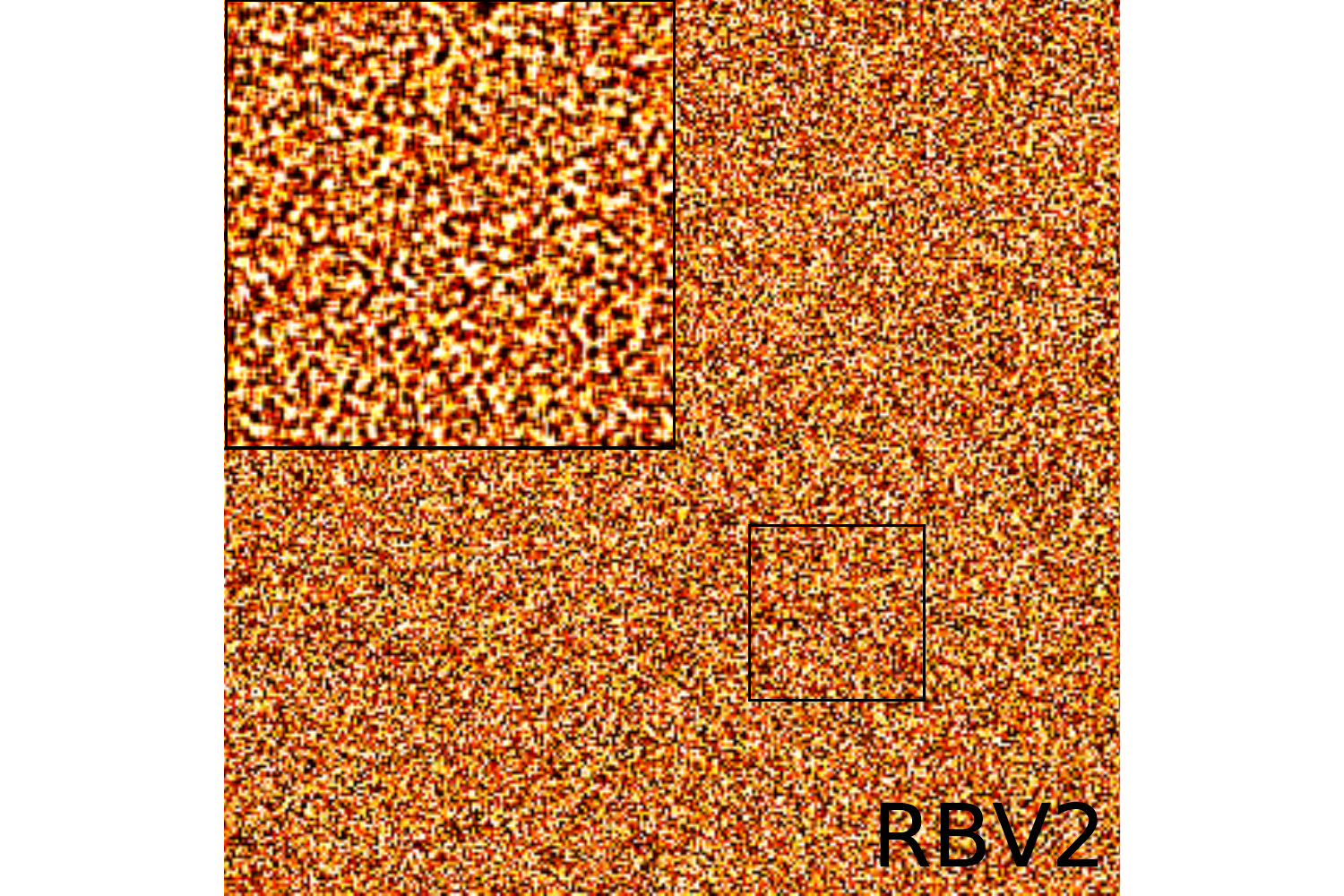}}
\resizebox{0.22\textwidth}{!}{\includegraphics*[trim={2.5cm .1cm 2.5cm .1cm},clip]{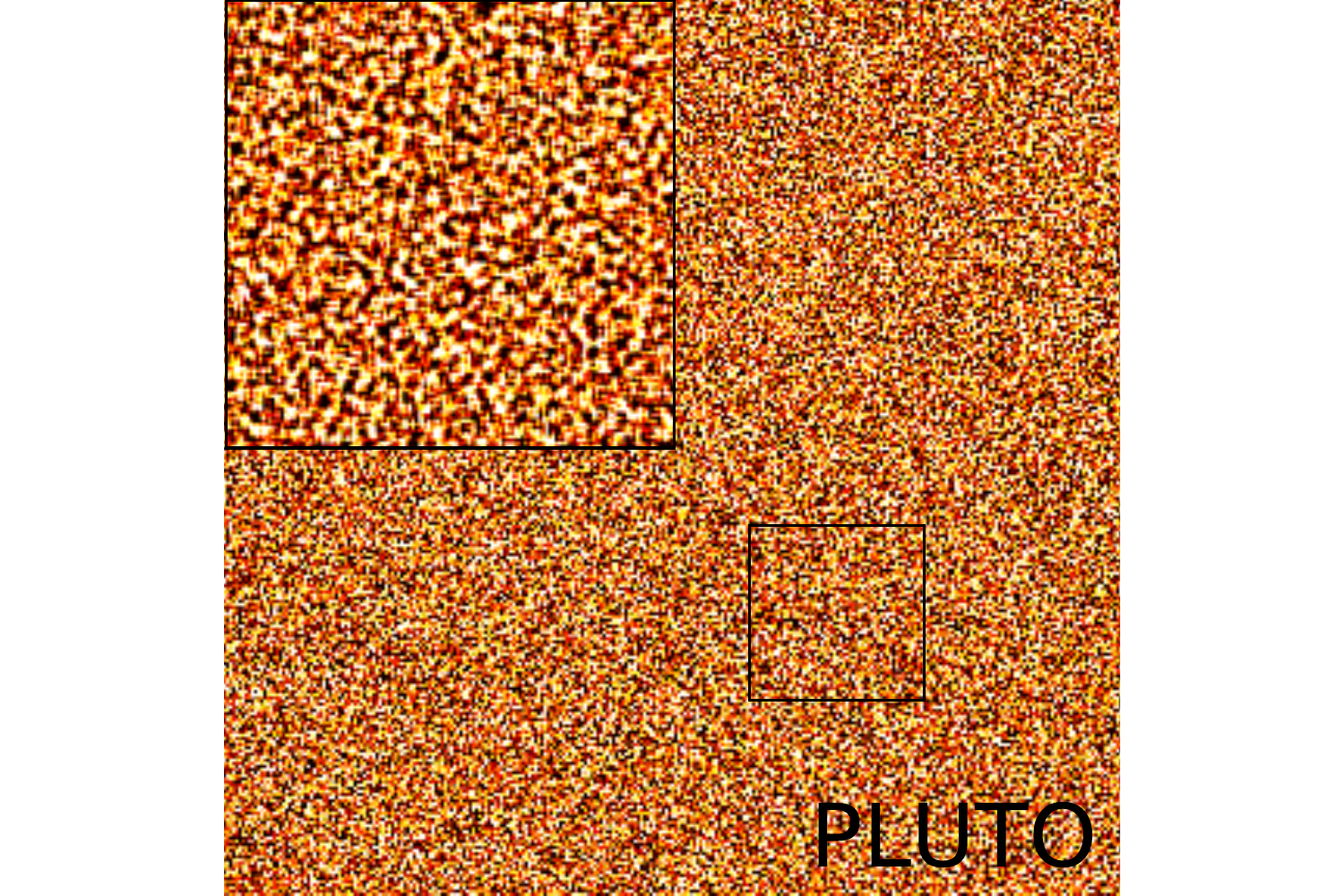}}
\resizebox{0.22\textwidth}{!}{\includegraphics*[trim={2.5cm .1cm 2.5cm .1cm},clip]{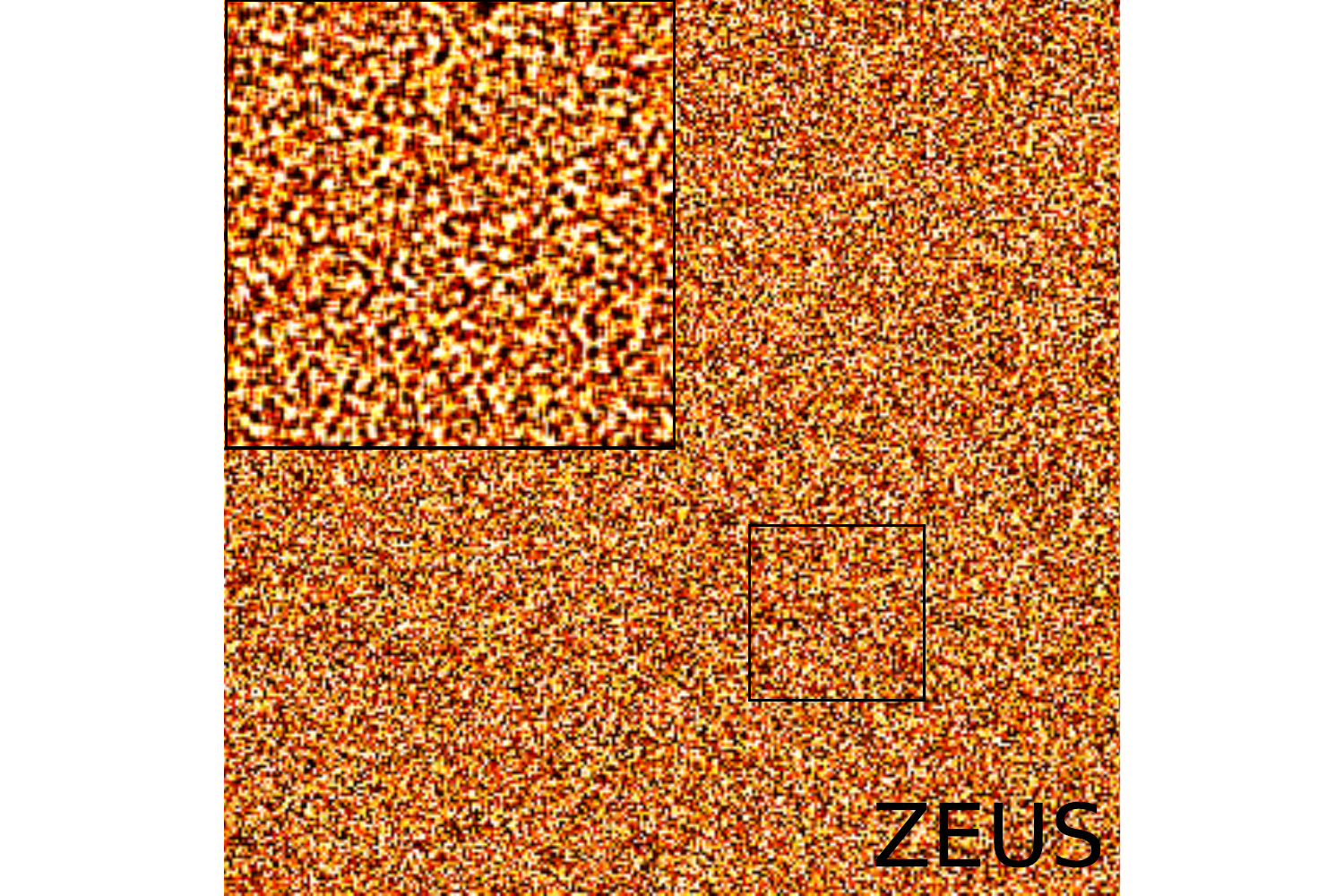}}
\resizebox{0.22\textwidth}{!}{\includegraphics*[trim={2.5cm .1cm 2.5cm .1cm},clip]{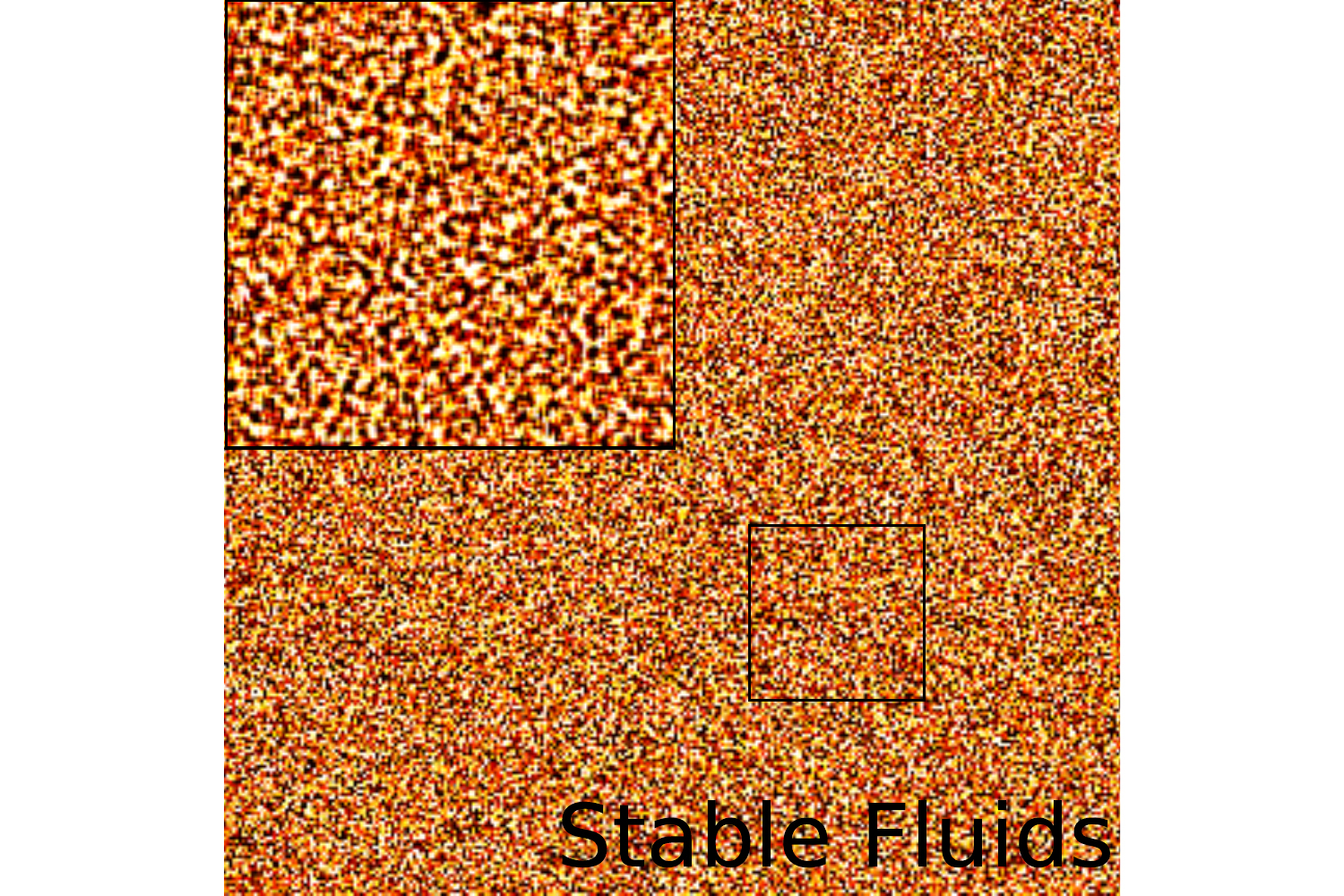}}
\resizebox{0.22\textwidth}{!}{\includegraphics*[trim={2.5cm .1cm 2.5cm .1cm},clip]{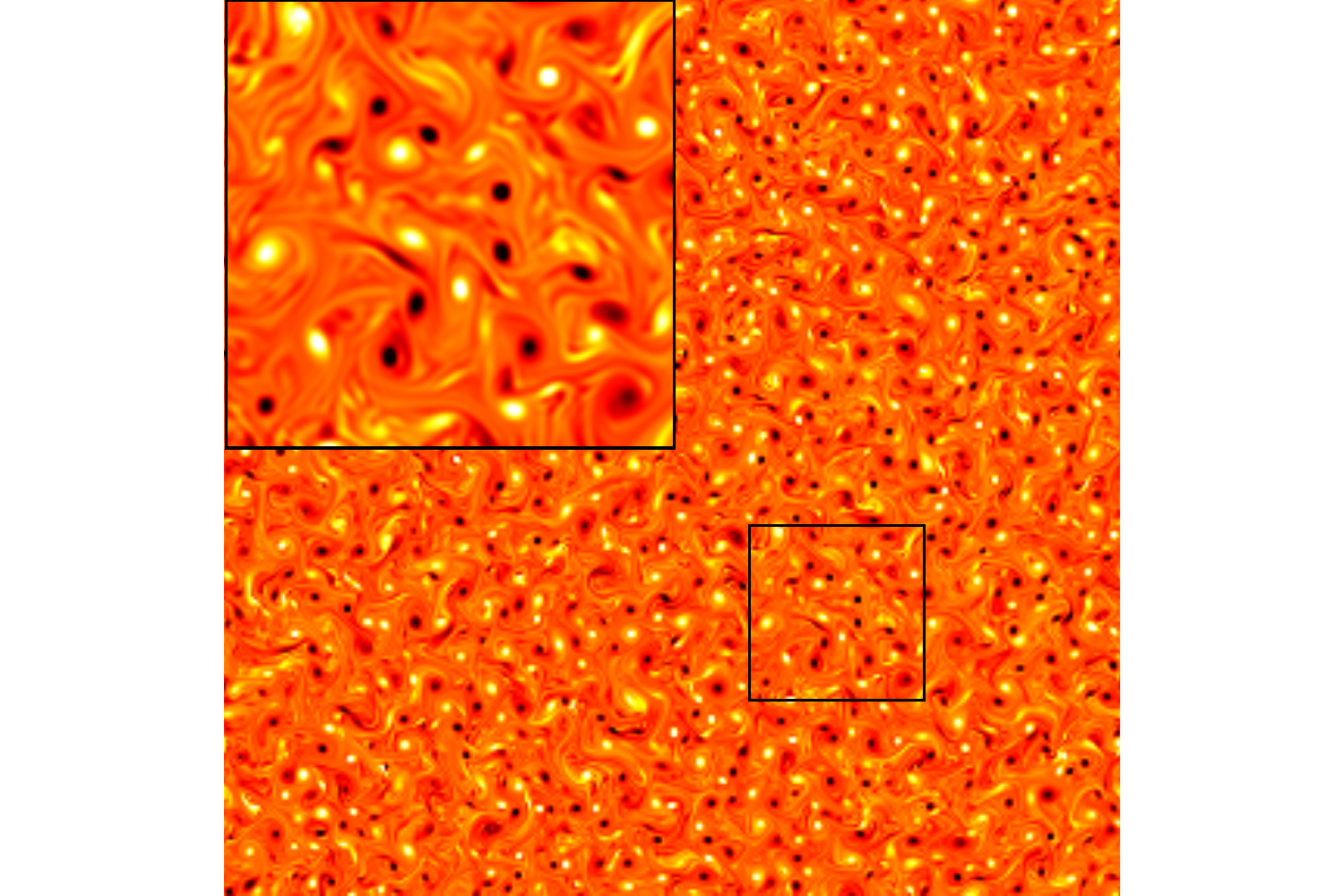}}
\resizebox{0.22\textwidth}{!}{\includegraphics*[trim={2.5cm .1cm 2.5cm .1cm},clip]{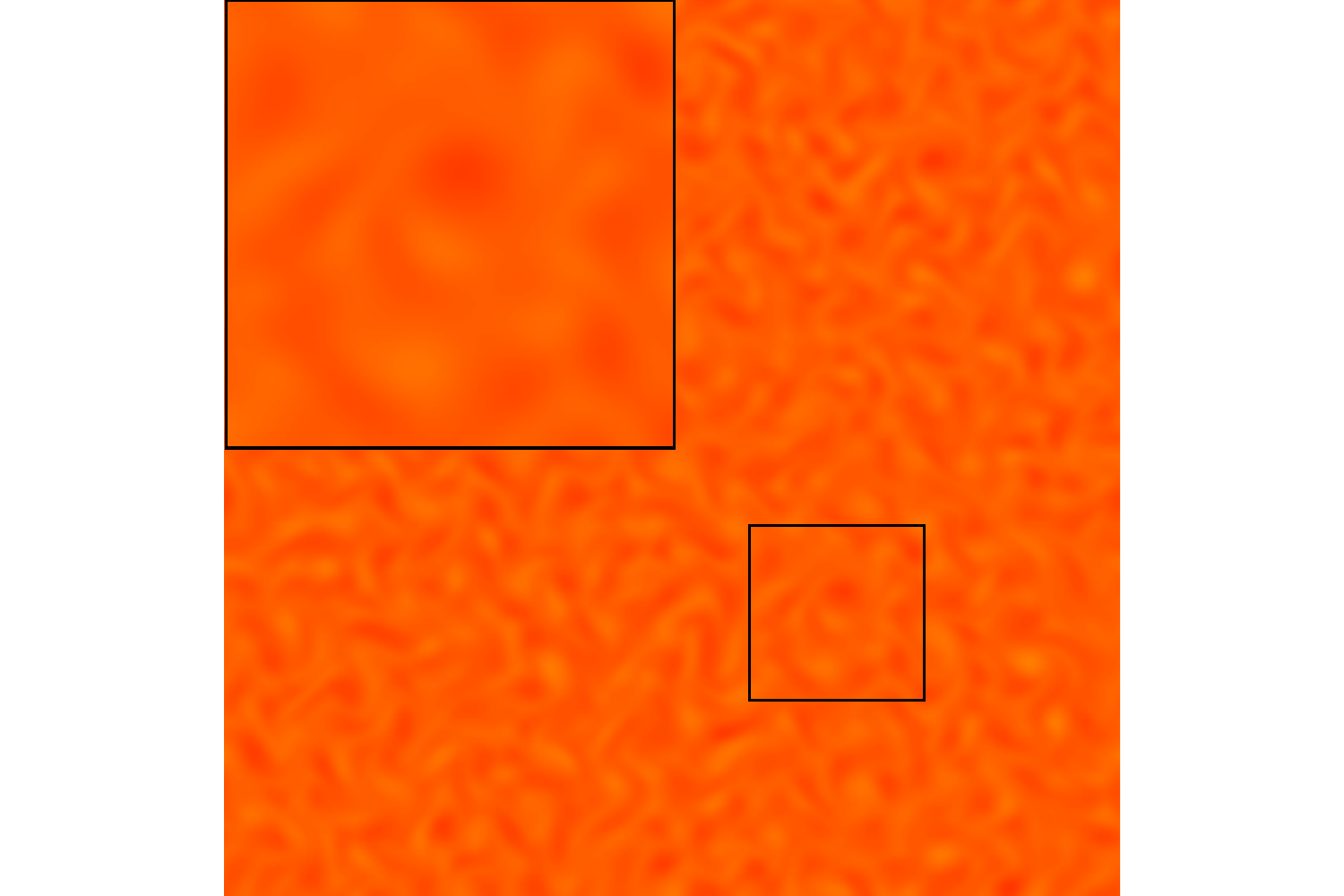}}
\resizebox{0.22\textwidth}{!}{\includegraphics*[trim={2.5cm .1cm 2.5cm .1cm},clip]{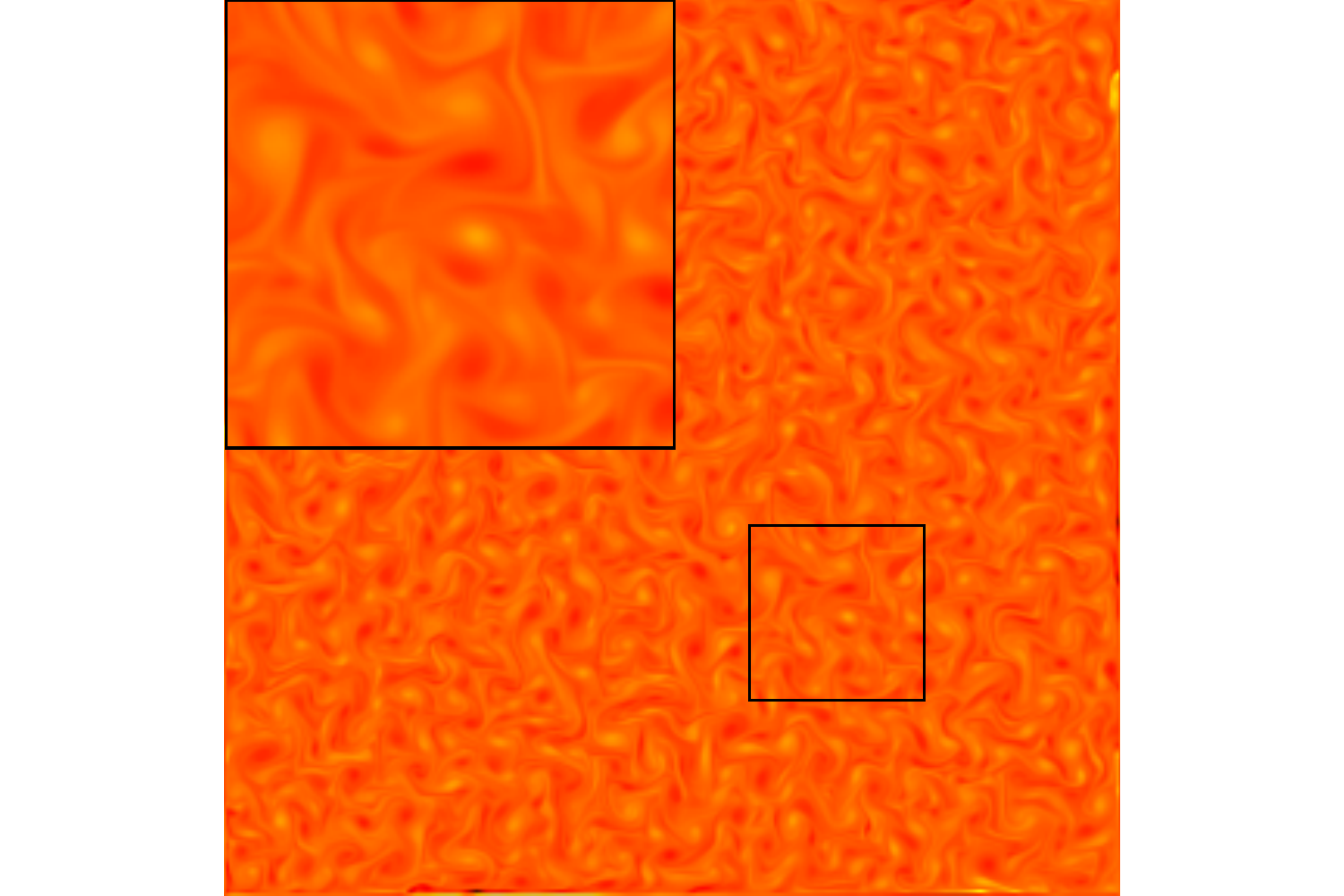}}
\resizebox{0.22\textwidth}{!}{\includegraphics*[trim={2.5cm .1cm 2.5cm .1cm},clip]{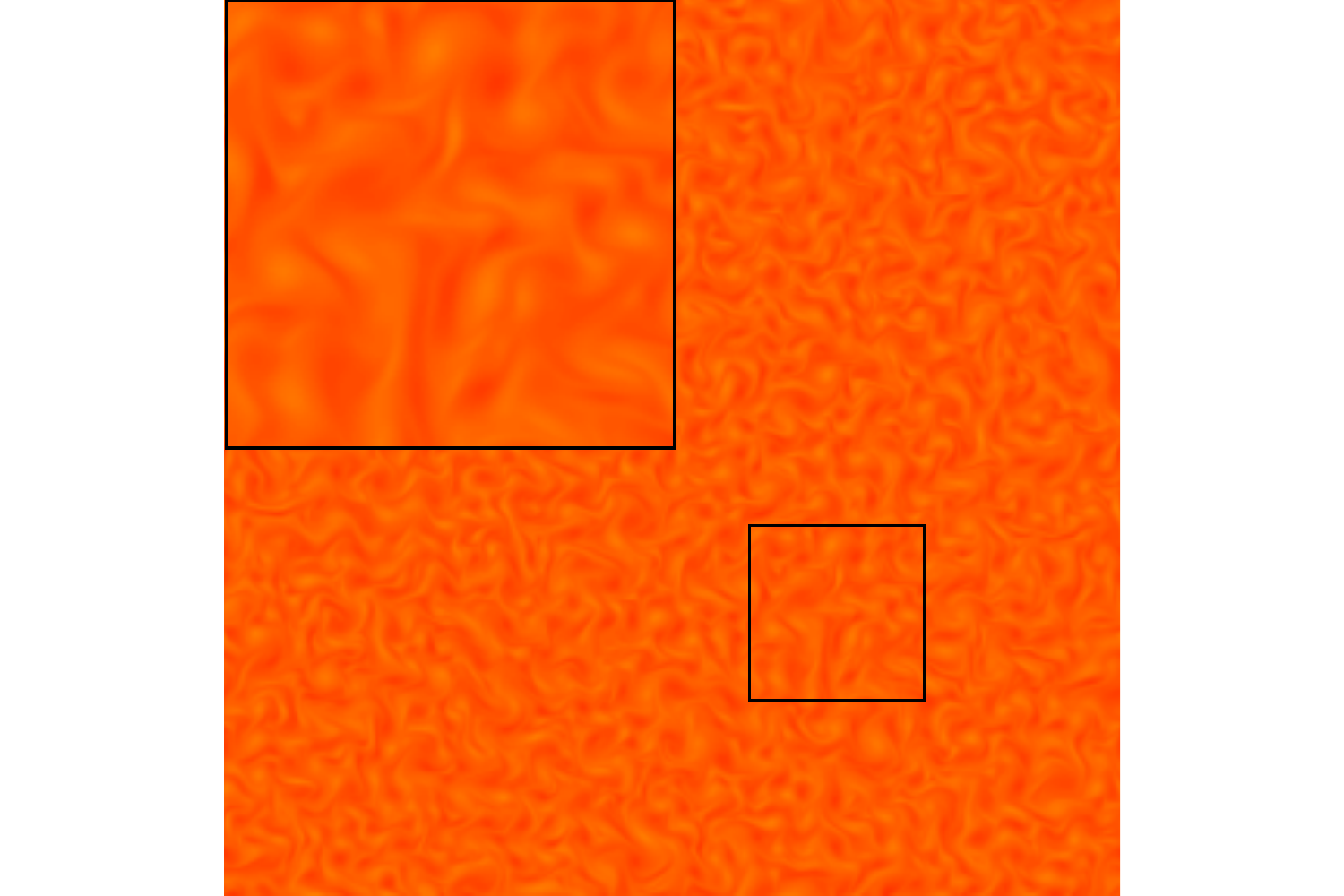}}
\resizebox{0.22\textwidth}{!}{\includegraphics*[trim={2.5cm .1cm 2.5cm .1cm},clip]{2dturb_vorticity_RBV2.pdf}}
\resizebox{0.22\textwidth}{!}{\includegraphics*[trim={2.5cm .1cm 2.5cm .1cm},clip]{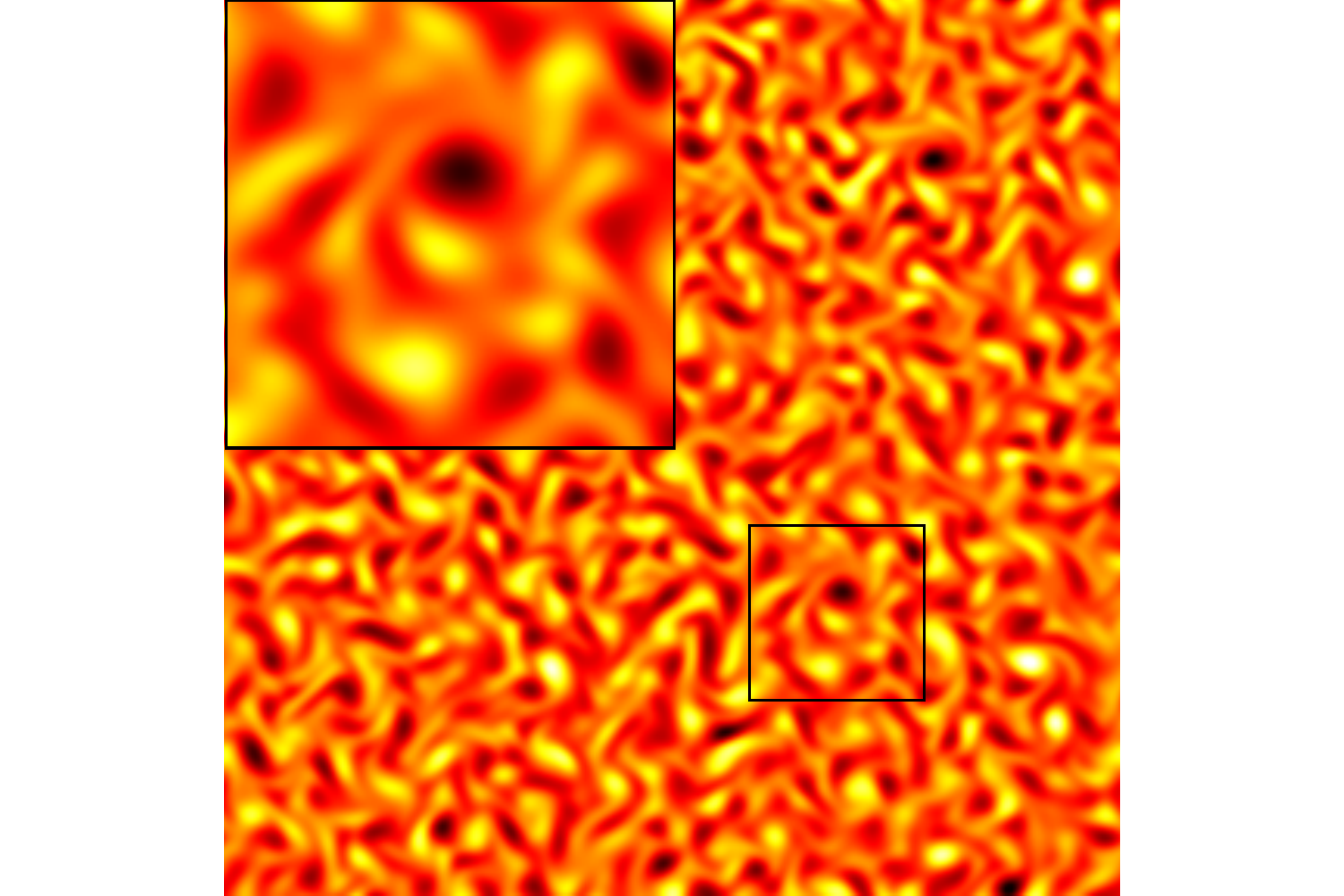}}
\resizebox{0.22\textwidth}{!}{\includegraphics*[trim={2.5cm .1cm 2.5cm .1cm},clip]{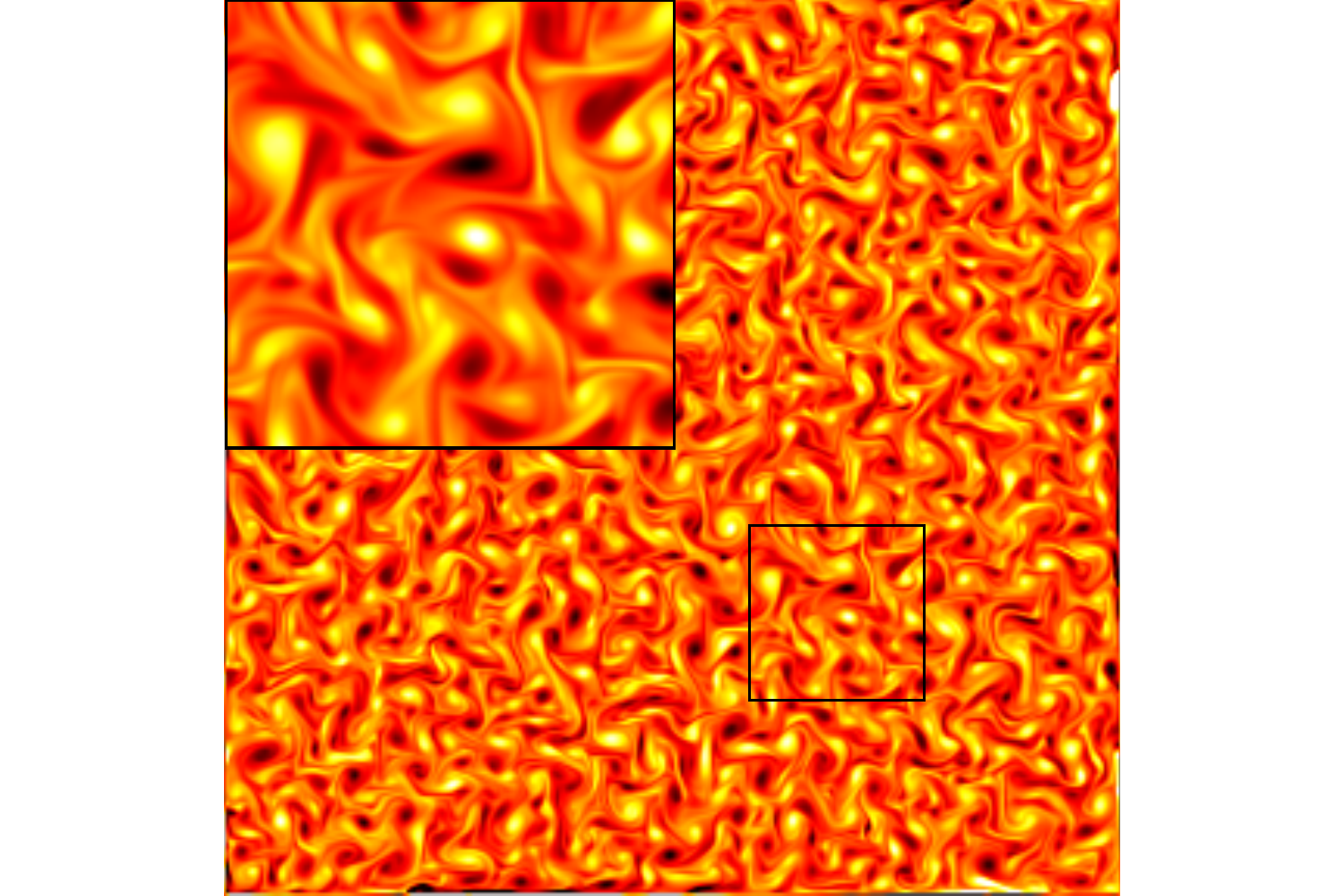}}
\resizebox{0.22\textwidth}{!}{\includegraphics*[trim={2.5cm .1cm 2.5cm .1cm},clip]{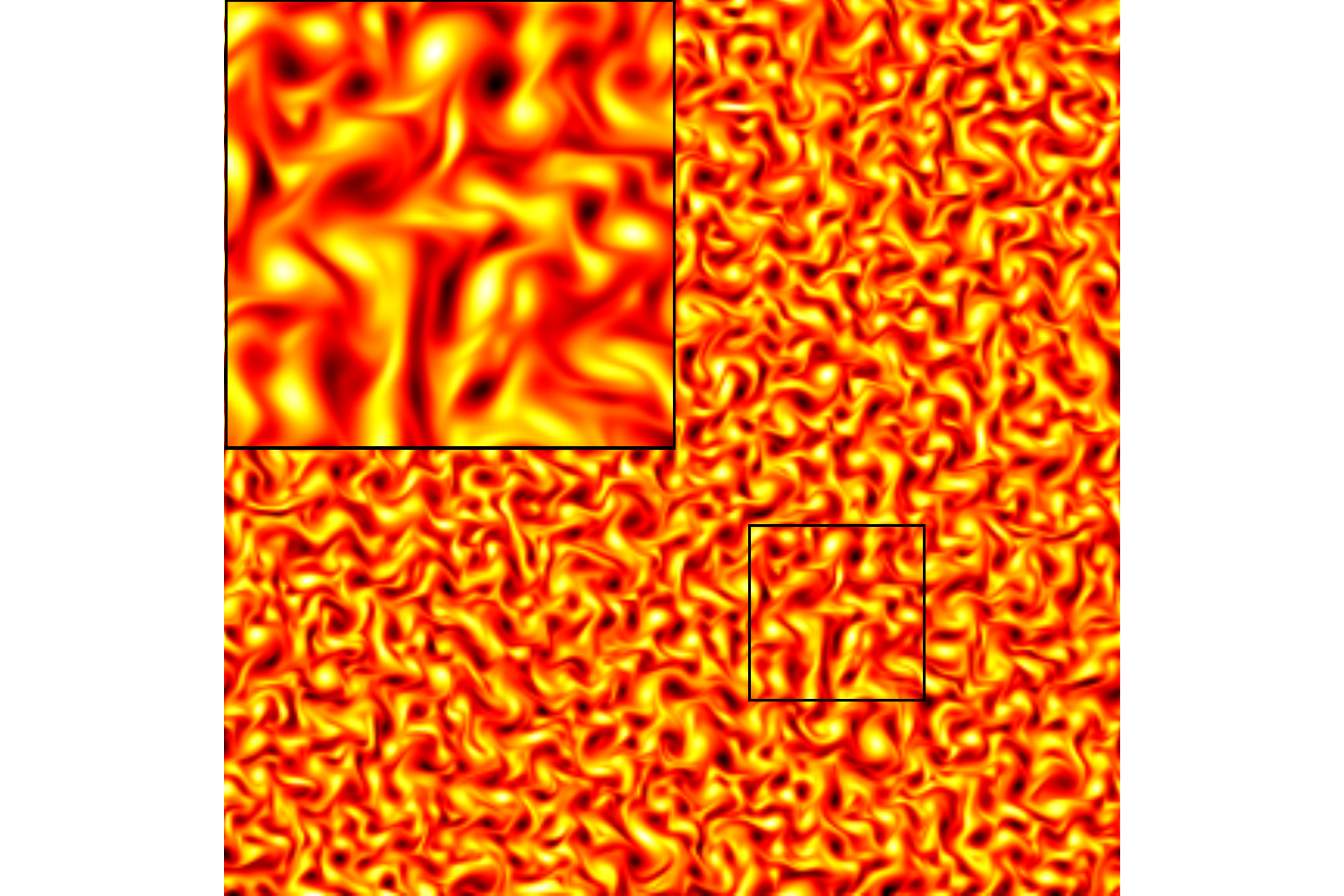}}

\end{center}
\caption{Vorticity evolution in  two-dimensional fluid simulations seeded with random, subsonic velocity fields smoothed with a low pass filter. Results are shown, in the columns running from left to right, for RBV2, PLUTO, ZEUS and Stable Fluids respectively, all on 1024x1024 zone grids. The initial vorticity conditions are displayed in the top row, and the middle and bottom rows show the results after $\sim 20$ sound crossing times of the simulation box. In the top and middle row, the vorticity color scale is held constant between $\pm.01$ for all eight  simulations. In the bottom row, the color scale is adjusted in PLUTO, ZEUS and Stable Fluids to $\pm-0.001$, $\pm-0.001$ and $\pm-0.002$ respectively, in order to show the evolution of the fluid vorticity structure. A magnification of a smaller portion of each simulation is shown in the upper left corner of each panel, corresponding to the region encapsulated by the smaller squares shown in the lower right of each panel. The simulations took 40 hours, 29 hours, 14 hours and 22 minutes for RBV2, PLUTO, ZEUS and Stable Fluids respectively.}
\label{fig:2D Turbulence Vorticity}
\end{figure*}

\begin{figure*}
\begin{center}
\resizebox{0.99\textwidth}{!}{\includegraphics*[trim={.15cm .1cm .25cm .01cm},clip]{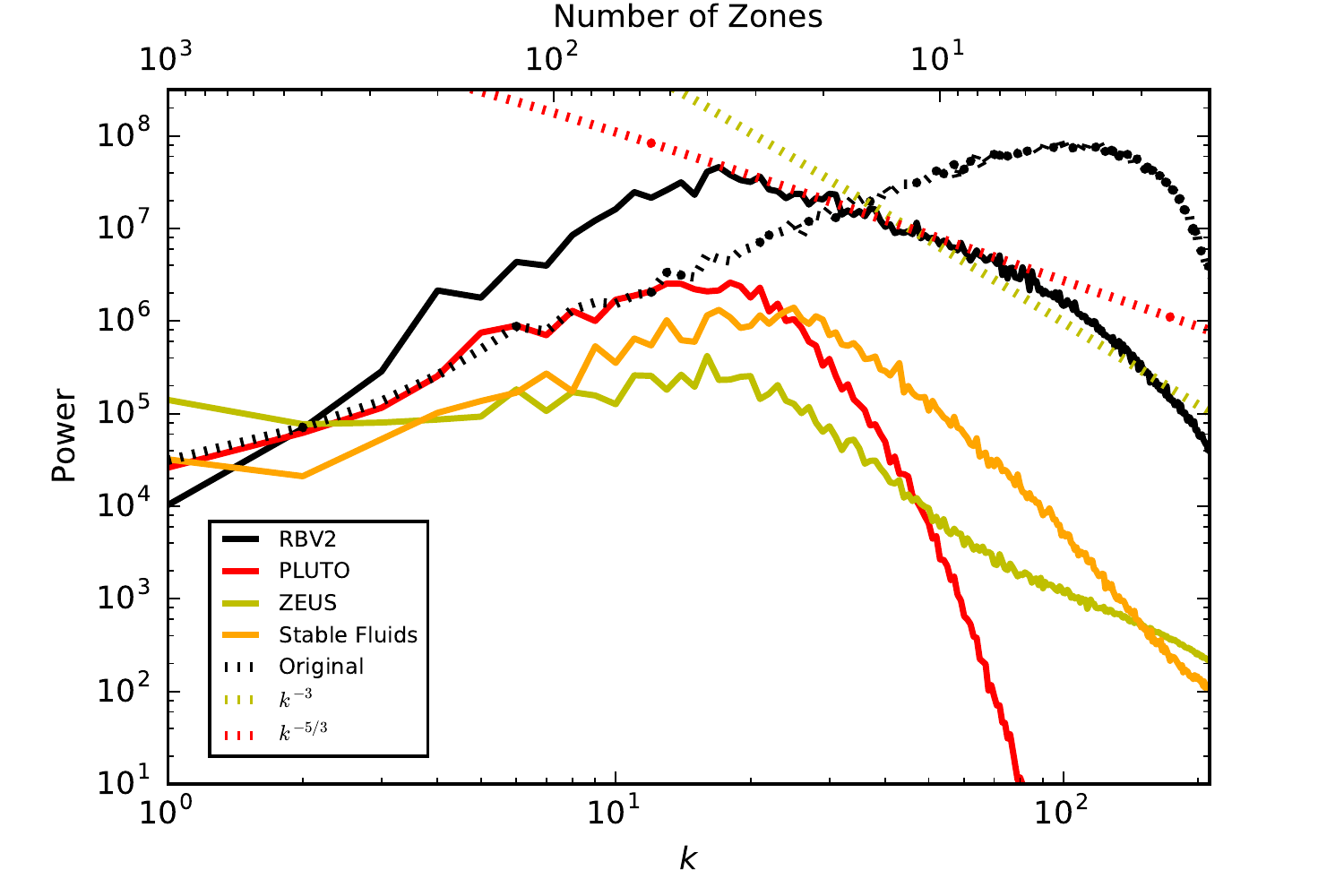}}
\end{center}
\caption{The resultant energy spectra (after 20 sound-crossing times of evolution) of the two-dimensional simulations depicted in Figure \ref{fig:2D Turbulence Vorticity}, for RBV2 (black solid), PLUTO (red solid), ZEUS (green solid) and Stable Fluids (yellow solid). Lines corresponding to spectra with slopes $k^{-5/3}$ and $k^{-3}$ are plotted with dashed orange and green lines respectively. The lower x-axis shows the wavenumber $k$ for each mode, while the upper x-axis shows the corresponding number of zones needed to represent one full cycle for each k disturbance.}
\label{fig:Power Spectrum}
\end{figure*}
We next present an initial assessment of the abilities of four of the numerical methods considered in this paper to develop the $k^{-3/2}$ and $k^{-3}$  power spectra associated with a two-dimensional turbulent cascade. More extensive tests in this area will be presented in a follow-up work (Seligman \& Laughlin 2017) that deals specifically with two-dimensional disk turbulence, and which explores the effects of both narrow and wide-band forcing. To make our initial test, we seeded a uniform initial density field with random, subsonic velocity fields smoothed with a low pass filter, and evolved the resulting initial condition with all four codes (RBV2, ZEUS, PLUTO, and Stable Fluids). The smoothing kernel effectively damps out small-scale velocity perturbations and creates an initial energy spectrum that lacks power in the highest frequency modes. 

With each code, we simulated flow on a 1024x1024 zone grid with identical sets of random x and y velocities in the range $-.1<v<.1$ (relative to a sound speed, $c_s=1$. We smoothed the initial field by averaging each velocity value with the 8 neighboring zones to remove small scale perturbations. We evolved the flows for $\tau \sim 20$ sound crossing times of the simulation box for all four numerical methods. Figure \ref{fig:2D Turbulence Vorticity} shows the vorticity in the initially seeded grid and the resulting end states of the simulations. The four columns, from left to right, show the evolution with RBV2, PLUTO, ZEUS and Stable Fluids. The color scale of the first row encompasses vorticities in the range $\pm.01$. The middle row shows the vorticity in the evolved flows with the same scaling, $\pm.01$. The last row shows the same evolved fields, with the scaling set to $\pm-0.001$, $\pm-0.001$ and $\pm-0.002$ in PLUTO, ZEUS and Stable Fluids, respectively. The revised color scales enhance the contrast in the decaying substructure that developed in those simulations. The simulations took 40 hours, 29 hours, 14 hours and 22 minutes for RBV2, PLUTO, ZEUS and Stable Fluids respectively.\footnote{Our initial prototype implementation of RBV2 can likely be further optimized for efficient performance.}

We calculate the energy spectrum by first taking the two-dimensional Fourier transforms of the $x$ and $y$ velocity fields and calculating
\begin{equation}\label{eq:powerspec}
    kE(k)=\frac{d E}{d\ln k}=k(|\hat{U}(k_x,k_y)|^{2}+|\hat{V}(k_x,k_y)|^{2}) \, ,
\end{equation}
where $k=2\pi/\lambda=\sqrt{k_x^2+k_y^2}$ is the two-dimensional wavenumber of the Fourier mode. Figure \ref{fig:Power Spectrum} shows the initial and time-evolved  energy spectra for such a setup with RBV2, PLUTO, ZEUS and Stable Fluids. 

The differences between the evolutionary sequences are profound.  RBV2 is able to maintain power in the high wavenumber range, while turbulent simulations in ZEUS and PLUTO clearly suffer from severe vortex dissipation. We expect that similar zeroth-order differences will also arise in turbulent simulations in which energy is continuously injected at scale, $k$. In a follow-up paper, we plan to present much higher resolution and longer simulations to demonstrate RBV2's capability to establish a true two-dimensional dual cascade. 
\subsection{Shock Tube}

\begin{figure}
\begin{center}
\resizebox{0.49\textwidth}{!}{\includegraphics*[trim={.25cm .1cm .25cm .1cm},clip]{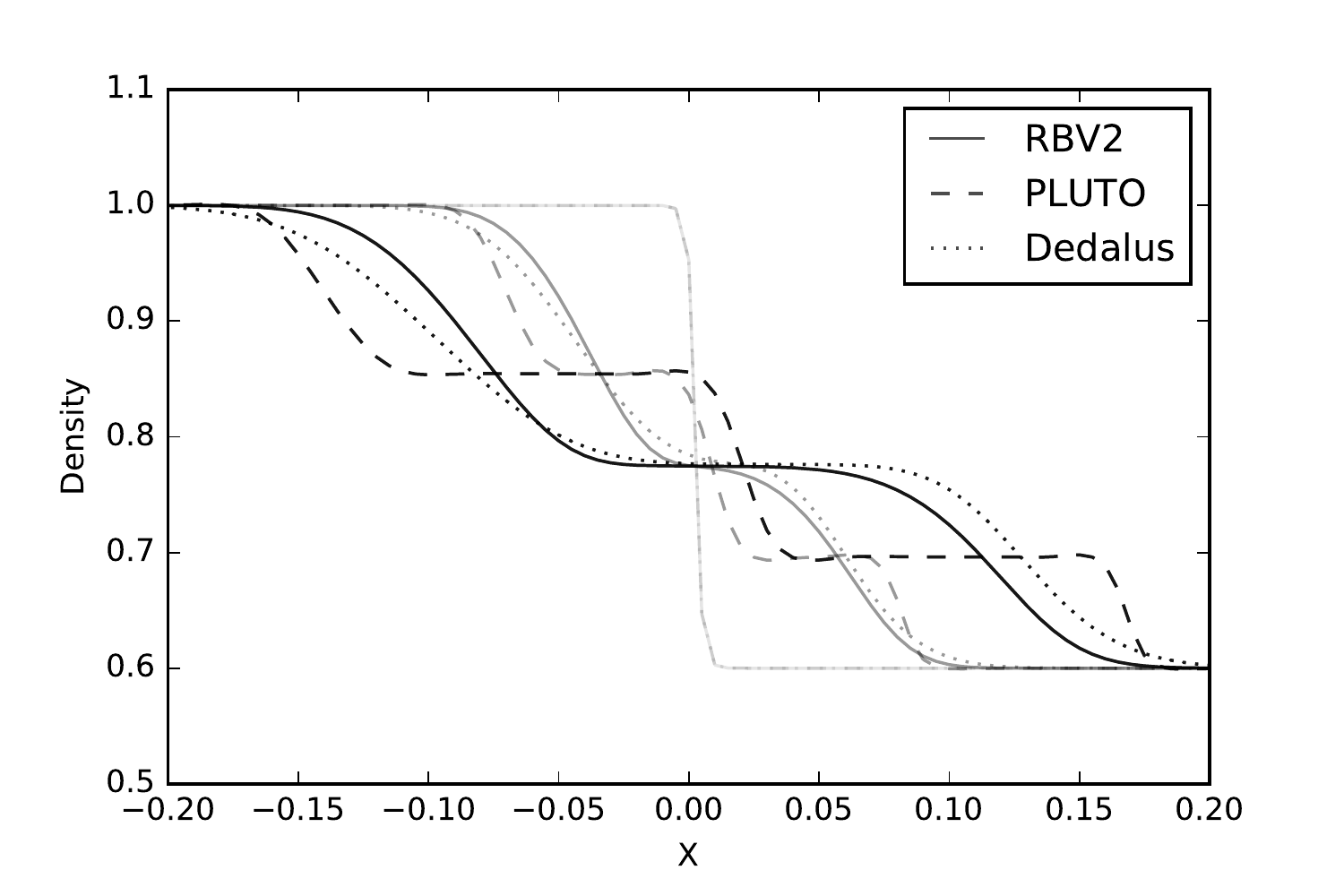}}
\caption{Simulations of a two-dimensional, isothermal shock tube with RBV2 (solid lines), PLUTO (dashed lines) and Dedalus (dotted lines). Decreasing transparency indicates more advanced simulation time, plotted at $0.00$, $0.01$ and $0.02$ sound crossing times. The computational domain is a 200x200 mesh that spans $(-.5,.5)$ and $(-.5,.5)$ with an initial density contrast of $.6$ and a sound speed $c_s=1.0$.}  
\label{fig:ShockTube}
\end{center}
\end{figure}

As a final test of the capabilities of RBV2, we test the code's ability to handle an isothermal shock tube. On a 200x200 mesh that spans $(-.5,.5)$ and $(-.5,.5)$, we simulate an isothermal shock tube with a density contrast of $.6$ and a sound speed $c_s=1.0$ with RBV2, Dedalus and PLUTO, shown in Figure \ref{fig:ShockTube}.

It is evident that both the current implementation of RBV2 and Dedalus are unable to capture the shock front with the same fidelity that PLUTO can. We conclude that the current implementation of RBV2 is most applicable in a subsonic regime of compressible fluid flow, and that in the instance of sonic or supersonic turbulent flows, an improved shock-handling algorithm should be implemented as part of the overall scheme.

\section{Simulations of vortex propagation in Keplerian Flow}
\subsection{The Shearing Box Approximation}

The Cartesian shearing box model \citep{Hawley1995} is frequently used to study local phenomenon in disks at a scale which captures the Keplerian shear, while omitting the global curvilinear geometry. The approximation is based on a local expansion of the hydrodynamical equations of motion at a fiducial disk radius $R_0$,

\begin{equation}\label{eq:SheerSheetEOM}
\frac{\partial {\bf v}}{\partial t} + {\bf v}\cdot\vec{\nabla} {\bf v}=-\frac{\vec{\nabla} P}{\rho}-2\boldsymbol{\Omega}\times {\bf v} +2q|\boldsymbol{\Omega}|^2 x \hat{x} \, .
\end{equation}

The term $2q|\boldsymbol{\Omega}|^2 x \hat{x}$ is the tidal expansion of the effective centrifugal and gravitational potential from the central object. For a Keplerian disk,
\begin{equation}
q=-\frac{d\ln|\boldsymbol{\Omega}|}{d\ln R}=\frac{3}{2}\, ,
\end{equation}
 and $-2{\bf \Omega}\times{\bf v}$ is the Coriolis acceleration. In the absence of a vertical gravitational force, a steady state solution to the equation is given by a flow with constant density, constant pressure and a uniform shear,

\begin{equation}\label{eq:SheerSheetSolution}
{\bf v}=-q|\boldsymbol{\Omega}| x \hat{y} \, .
\end{equation}

The shearing box approximation is completed by a description of the flow at the boundaries.  The computational domain is a rectilinear, with sides $L_x$ and $L_y$, on a mesh that spans $(-L_x/2,L_x/2)$ and $(-L_y/2,L_y/2)$. The box is assumed to be surrounded by identical shearing boxes, so the boundaries are strictly periodic at $t=0$. As the flow evolves and the patch co-rotates with the Keplerian shear at the fiducial radius, the azimuthal boundaries remain  periodic. However, the radial boundary conditions are determined by the location of the neighboring boxes, since they have sheared with respect to the fiducial radius. These boundary conditions can be written as,
\begin{equation}\label{eq:SheerSheetRadialBoundary}
f(x,y)=f(x\pm L_x,y\mp (q\boldsymbol{\Omega}L_x t)\text{mod} L_y)  \, ,
\end{equation}
and
\begin{equation}\label{eq:SheerSheetAzimuthalBoundary}
f(x,y)=f(x,y\pm L_y)  \, ,
\end{equation}
for the radial and azimuthal boundaries respectively.

These boundary conditions must be modified for the azimuthal velocity and momentum at the radial boundaries to account for the underlying differential shear across the simulation domain, such that

\begin{equation}\label{eq:SheerSheetRadialBoundary}
v_y(x,y)=v_y(x\pm L_x,y\mp (q\boldsymbol{\Omega}L_x t)\text{mod} L_y) \pm q\boldsymbol{\Omega}L_x \, ,
\end{equation}
and
\begin{equation}\label{eq:SheerSheetRadialBoundary}
s_y(x,y)=s_y(x\pm L_x,y\mp (q\boldsymbol{\Omega}L_x t)\text{mod} L_y) \pm q\boldsymbol{\Omega}L_x\rho \, .
\end{equation}
At any given timestep in the simulation, the image of a radial ghost zone will generally lie between two mesh zones, so we use a first order interpolation to find the appropriate value.

In theory, the shearing sheet model with its ``semi-periodic'' boundary conditions is amenable to higher order filtering as described by \citet{lele1992}. One could imagine constructing filter kernels at each time step such that the ghost zones were convolved with the correctly interpolated image values. We found that \citet{lele1992}'s high order filter with non-periodic boundaries sufficiently damped high frequency perturbations everywhere except at the azimuthal boundaries. Not only are large velocity gradients continuously injected here, but the Lele filter is also not optimized for non-periodic boundaries. 

Since the matrix inversion necessary for the filter is computationally intensive, we implement a stochastic soaking method in addition to the purely non-periodic filter for all shearing box simulations presented with RBV2. At a cadence that corresponds to roughly once every sound crossing time for the simulation domain, we replace fluid quantities in the edge zones with a mixture of the calculated flow quantities and the analytic shear solution, such that the outer-most zone is entirely replaced with the analytic shear solution, and each inner zone is replaced with a mixture of the calculated solution and the analytic shear until the inner-most soak zone carries the full calculated solution. At each timestep when the soaking is required, we randomly assign a number of soak zones that correspond to an edge soak domain of at most fifteen percent of the radial extent. This method allows disturbances to propagate out of the simulation domain without exhibiting spurious reflection, and prevents numerical instabilities that would otherwise manifest at the boundaries from propagating into the domain of interest.

\subsection{Vortex in a Keplerian Shear Flow}
To quote \citet{Abramowicz1992}, ``every rotating cosmic fluid that can be observed sufficiently closely displays either vortices or magnetic flux tubes on it surface; examples are tornadoes in the Earth's atmosphere, the Great Red Spot and other vortices in Jupiter's atmosphere, and sunspots.'' \citet{Abramowicz1991} suggest that accretion disks also harbor dynamically significant vortices, and presented an analytic argument for the interpretation of the power spectra of X-ray variability produced by a hot accretion disk replete with vortices and magnetic flux tubes. Young stellar objects also tend to exhibit strong temporal variability, and observations such as those presented by \citet{Flaherty2016} may be explained with a similar physical arguments. 

In addition to the indirect observational clues of presence, gaseous vortices in protostellar disks are potentially attractive sites for planet formation \citep{barge1995,Tanga1996}. \citet{barranco2005}, using three-dimensional simulations, found that off-midplane vortices form naturally and are stable for many  hundreds of orbits.  Due to the rapid settling of particles into their central regions, and the attendant enhancement of the local surface density, vortices in discs have been suggested as triggers of giant planet formation via gravitational instability \citep{adams1995}. \citet{Hodgson1998} found no overdensity of particles in full three-dimensional simulations of turbulent disks, although they admitted that the vortices in their simulations were very short lived.

\begin{figure}
\begin{center}
\resizebox{0.49\textwidth}{!}{\includegraphics*[trim={2.5cm .1cm 2.5cm .1cm},clip]{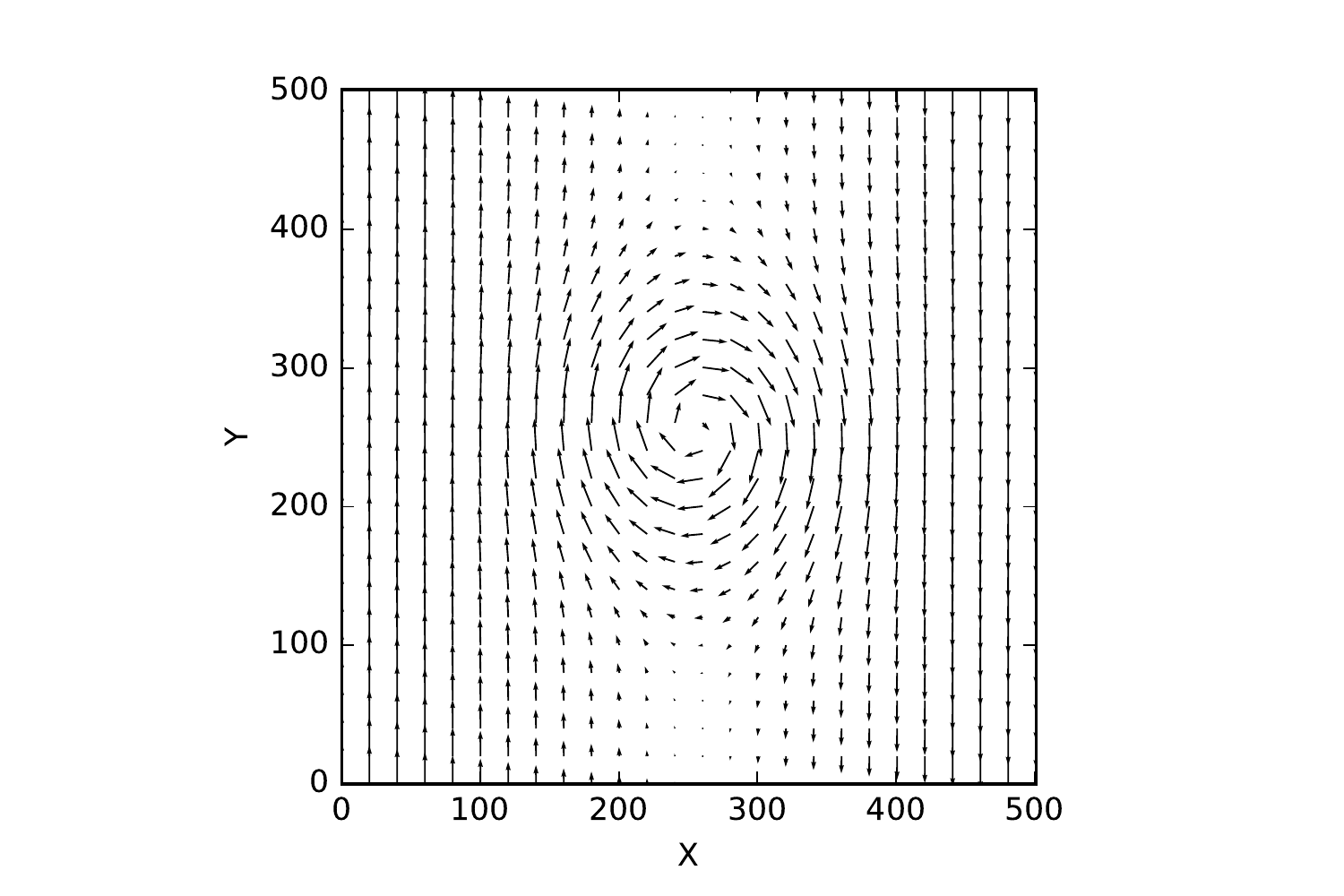}}
\caption{Initial  velocity  profile  of  the  vortex  seeded  into a Keplerian shear flow, evolved by the RBV2, PLUTO, and ZEUS algorithms for compressible flow, and depicted in Figure \ref{fig:VortexDiskVorticity}. The vortex profile is created following the prescription given by \citet{adams1995}. We use shearing box simulations \citep{Hawley1995} with a radial and azimuthal extent of $\Delta R=.04$~AU centered at $R=1$~AU with a sound speed, $c_s=0.050$, such that the Mach number of the shear at the radial boundaries, $v=\pm.029$, is ${\cal M}\approx0.75$.  The maximum vortex velocity, $v=\pm.025$, is comparable to the maximum shear velocity, such that close to the fiducial radius, the flow is dominated by the vortex. Note that the units on the X and Y axis indicate the grid elements of the simulation.}
\label{fig:VortexDiskVelocity}
\end{center}
\end{figure}

Motivated by the possibility of future investigations of disk vortices with RBV2, we run shearing box numerical simulations of an accretion disk centered at  $R=1$~AU from the central object with RBV2, PLUTO and ZEUS. We simulate a radial and azimuthal extent of $\Delta R=.04$~AU, centered around the fiducial radius, with maximum shear at the radial boundaries, $v=\pm.029$, on 512x512 zones. We seed a vortex (Figure \ref{fig:VortexDiskVelocity})
 following the prescription of \citet{adams1995}, multiplied by an exponential function to damp the vortex at the boundaries. Specifically, with  $k_RR=\frac15$, for small radii, $0<r <R$, the velocity perturbation is given by
\begin{figure}
\begin{center}
\resizebox{0.375\textwidth}{!}{\includegraphics*[trim={3.5cm 1.cm 3.5cm 1.cm},clip]{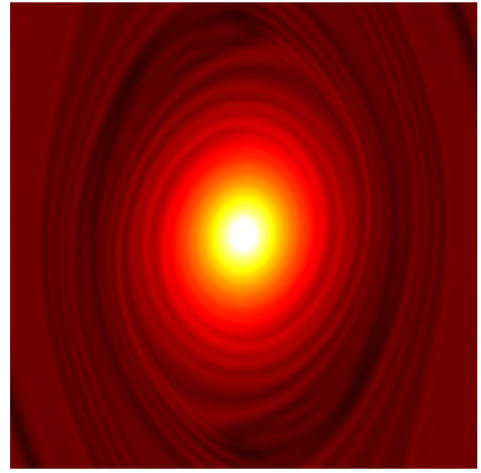}}
\resizebox{0.375\textwidth}{!}{\includegraphics*[trim={3.5cm 1.cm 3.5cm 1.cm},clip]{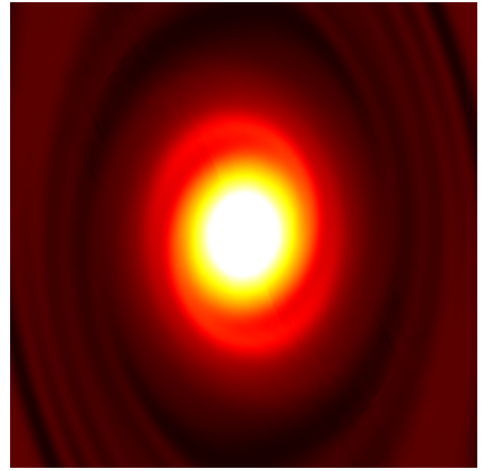}}
\resizebox{0.375\textwidth}{!}{\includegraphics*[trim={3.5cm 1.cm 3.5cm 1.cm},clip]{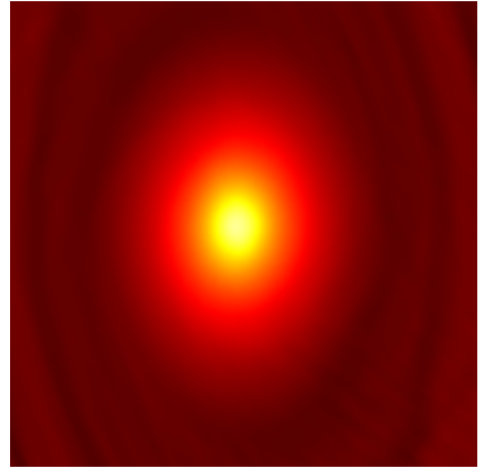}}
\end{center}
\caption{Evolved structure of a vortical disturbance seeded into a Keplerian shear flow, see Figure \ref{fig:VortexDiskVelocity}, evolved for $\sim15$ vortex sound crossing times with RBV2 (top), PLUTO (middle) and ZEUS (bottom). Since the seeded vortex is not in an analytic equilibrium, it is expected that the vortex should shed Rossby waves. It appears that the substructure that develops in the flow around the vortex due to the shedding of these waves is very different in all three cases.}
\label{fig:VortexDiskVorticity}
\end{figure}

\begin{equation}\label{eq:FredVort1}
    {\bf v}=\frac{\Gamma}{\pi R} K_1(k_R R)I_1(k_R r)\exp(-300r)\hat{\phi} \, ,
\end{equation}
and for $r>R$
\begin{equation}\label{eq:FredVort2}
    {\bf v}=\frac{\Gamma}{\pi R} I_1(k_R R)K_1(k_R r)\exp(-300r)\hat{\phi} \, ,
\end{equation}

where $K_1$ and $I_1$ are modified Bessel functions, $\Gamma$ is the vortex strength and $k_R$ is the Rossby wavenumber,

\begin{equation}\label{eq:Rossby}
    k_R=\frac{\Omega}{c_s} \, .
\end{equation}
For a derivation of the functional form of such vortices, see \citet{adams1995}. The sound speed, $c_s=.050$, set by the scale height of the disk, is such that the Mach number of the shear at the radial boundaries, $v=\pm.039$, is ${\cal M}\approx0.75$.  The maximum vortex velocity, $v=\pm.025$, is comparable to the maximum shear velocity, such that close to the fiducial radius, the flow is dominated by the vortex.

Figure \ref{fig:VortexDiskVorticity} shows the vorticity in the simulations after roughly 15 vortex sound crossing times. Unlike the idealized vortex proposed by \citet{Yee1999}, these structures are not initially in an equilibrium state. The construction of an analytic vortex in equilibrium in the presence of a shear flow and a coriolis deflection has, as yet, eluded an exact solution. Therefore, we expect the vortex to shed Rossby waves. In examining the results of the parallel simulations, we find that the substructure that develops in the accretion disk flow around the vortex due to these waves is very different in all three cases, with the RBV2 solution being most centrally peaked, and displaying a more finely detailed flow pattern in the outer regions of the vortex

\subsection{Dynamical Interaction of Two Vortices in a Disk}

The dynamical interaction of vortices has received a great deal of attention, partly  because the vortex merger process drives the decay of and growth of coherent structure in two-dimensional turbulence \citep{Hopfinger1993} and may be partly responsible for the dynamics of three-dimensional turbulence \citep{Vincent1991}. In our context, the interaction and merger of vortices may have important consequences for determining the form of eddy viscosity manifested in thin astrophysical disks.

It is well known from laboratory experiments \citep{Griffiths1987} and numerical simulations \citep{Overman1982,Dritschel1985}   that vortices of the same spin tend to attract each other and may merge given proper vortex strengths and separation. \citet{Cerretelli2003} demonstrated that  the resulting anti-symmetric vorticity field causes the two vortices to be pushed together and that the number of vortex orbits preceding a merger increases with the effective Reynolds number of the flow.

In the context of protoplanetary disks, \citet{Godon1999} found that the merger and interaction of long-lived, anticyclonic vortices may play an important role in the formation process for giant planets. If vortices serve as sites for dust-particle settling, it seems plausible that the dynamical merger of two of these vortices may be important for the planet formation process.

\begin{figure}
\begin{center}
\resizebox{0.20\textwidth}{!}{\includegraphics*[trim={4.5cm 1.4cm 4.5cm 2.0cm},clip]{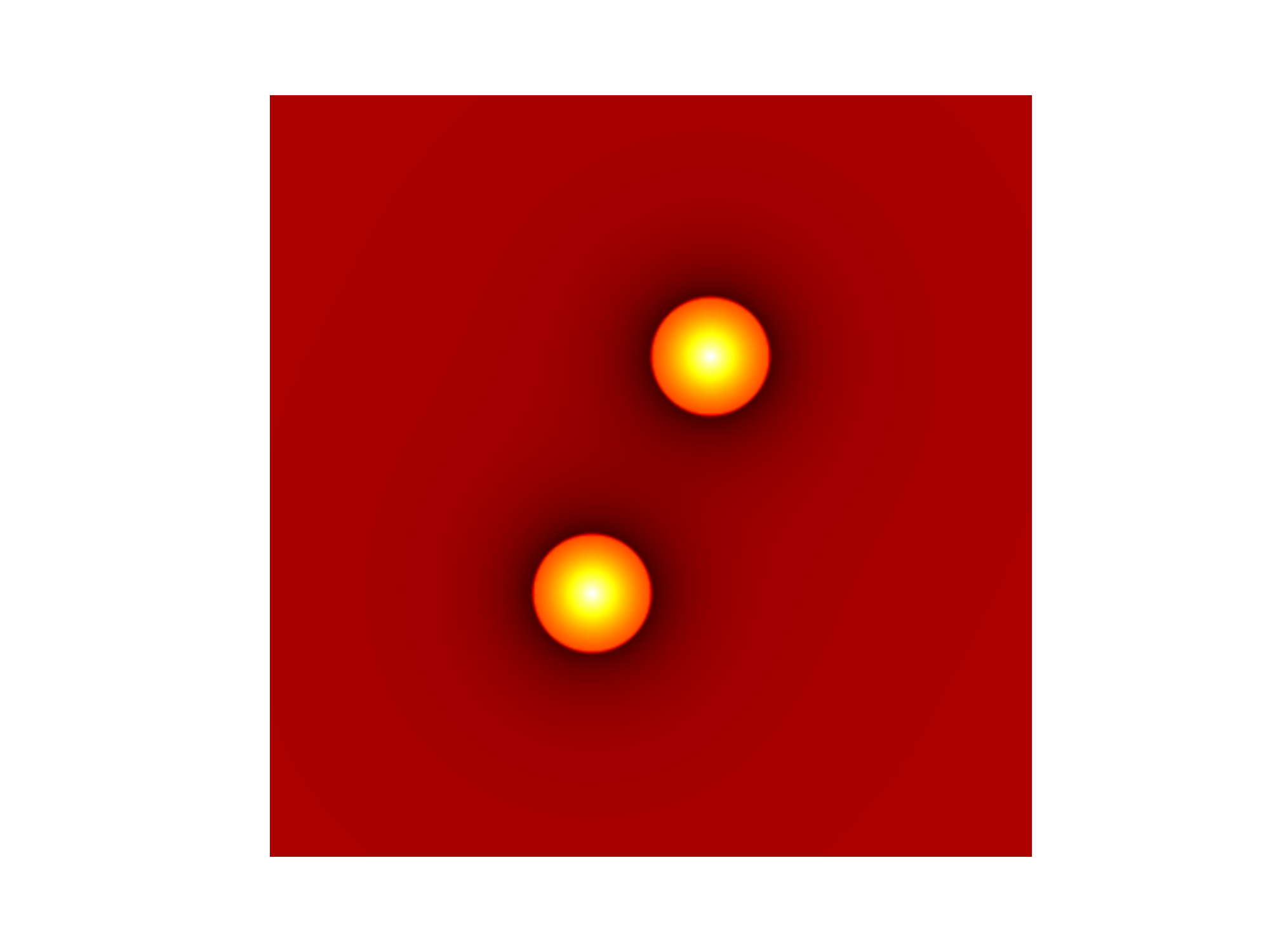}}
\resizebox{0.20\textwidth}{!}{\includegraphics*[trim={4.5cm 1.4cm 4.5cm 2.0cm},clip]{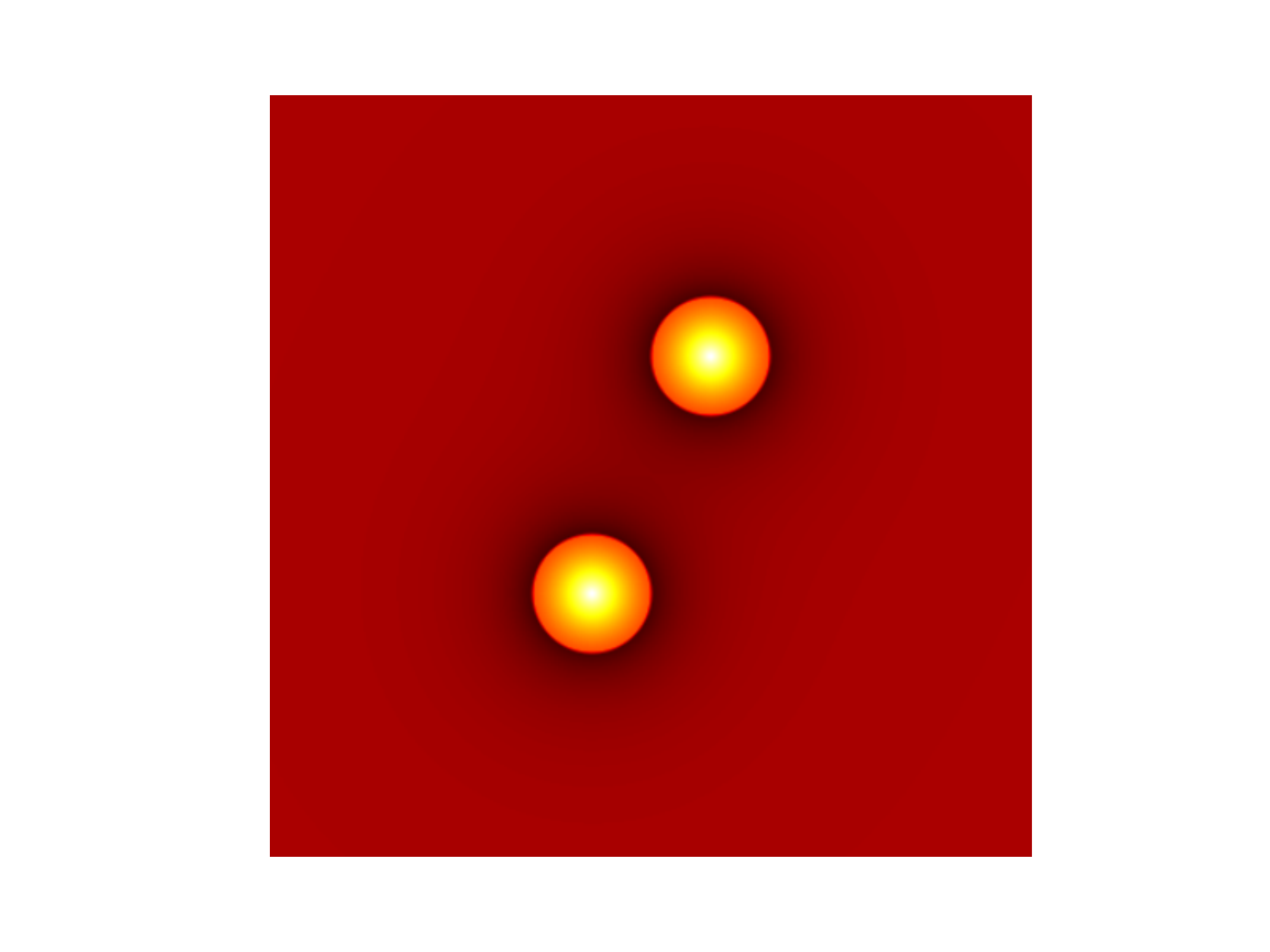}}
\resizebox{0.20\textwidth}{!}{\includegraphics*[trim={4.5cm 1.4cm 4.5cm 2.0cm},clip]{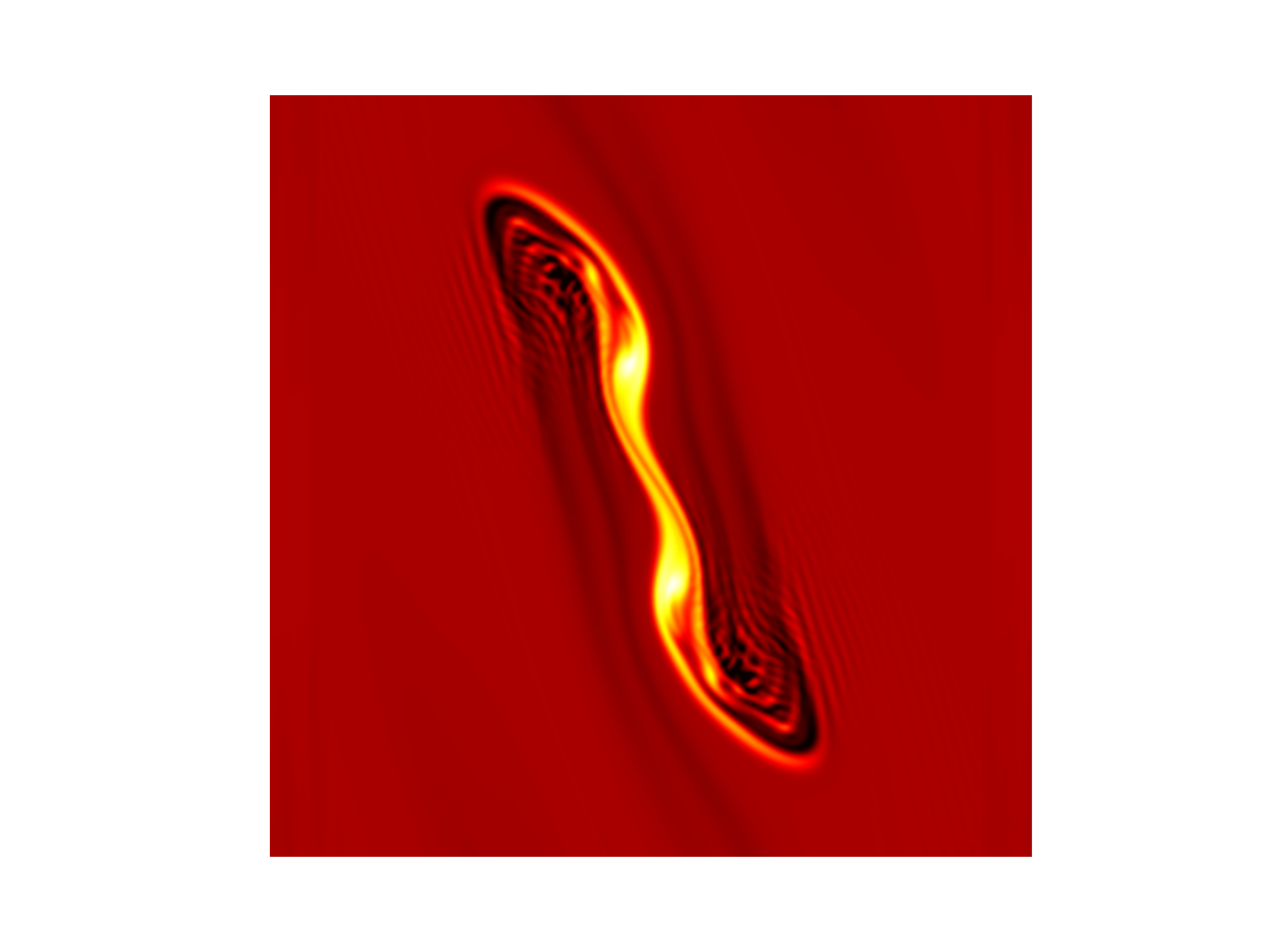}}
\resizebox{0.20\textwidth}{!}{\includegraphics*[trim={4.5cm 1.4cm 4.5cm 2.0cm},clip]{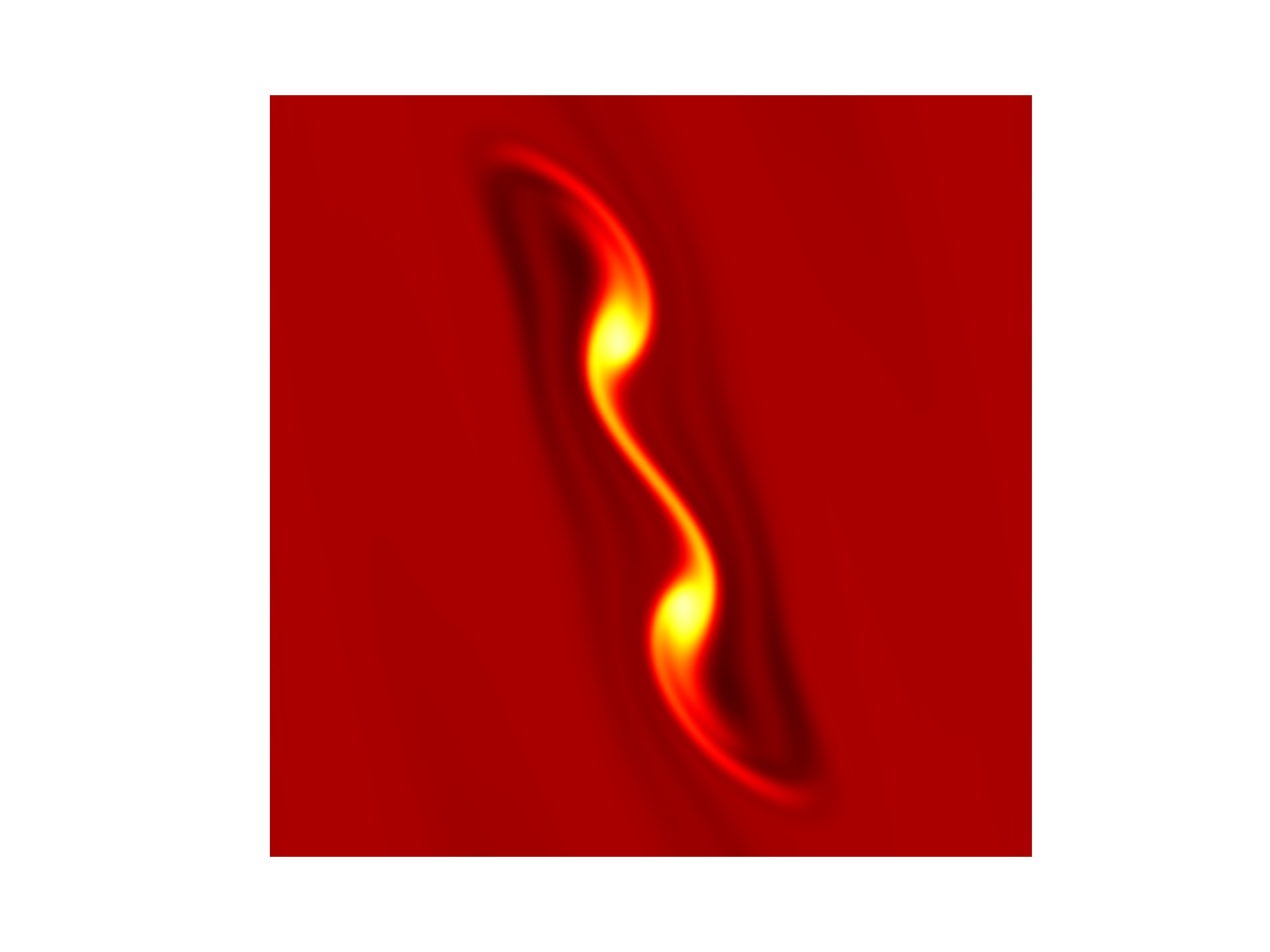}}
\resizebox{0.20\textwidth}{!}{\includegraphics*[trim={4.5cm 1.4cm 4.5cm 2.0cm},clip]{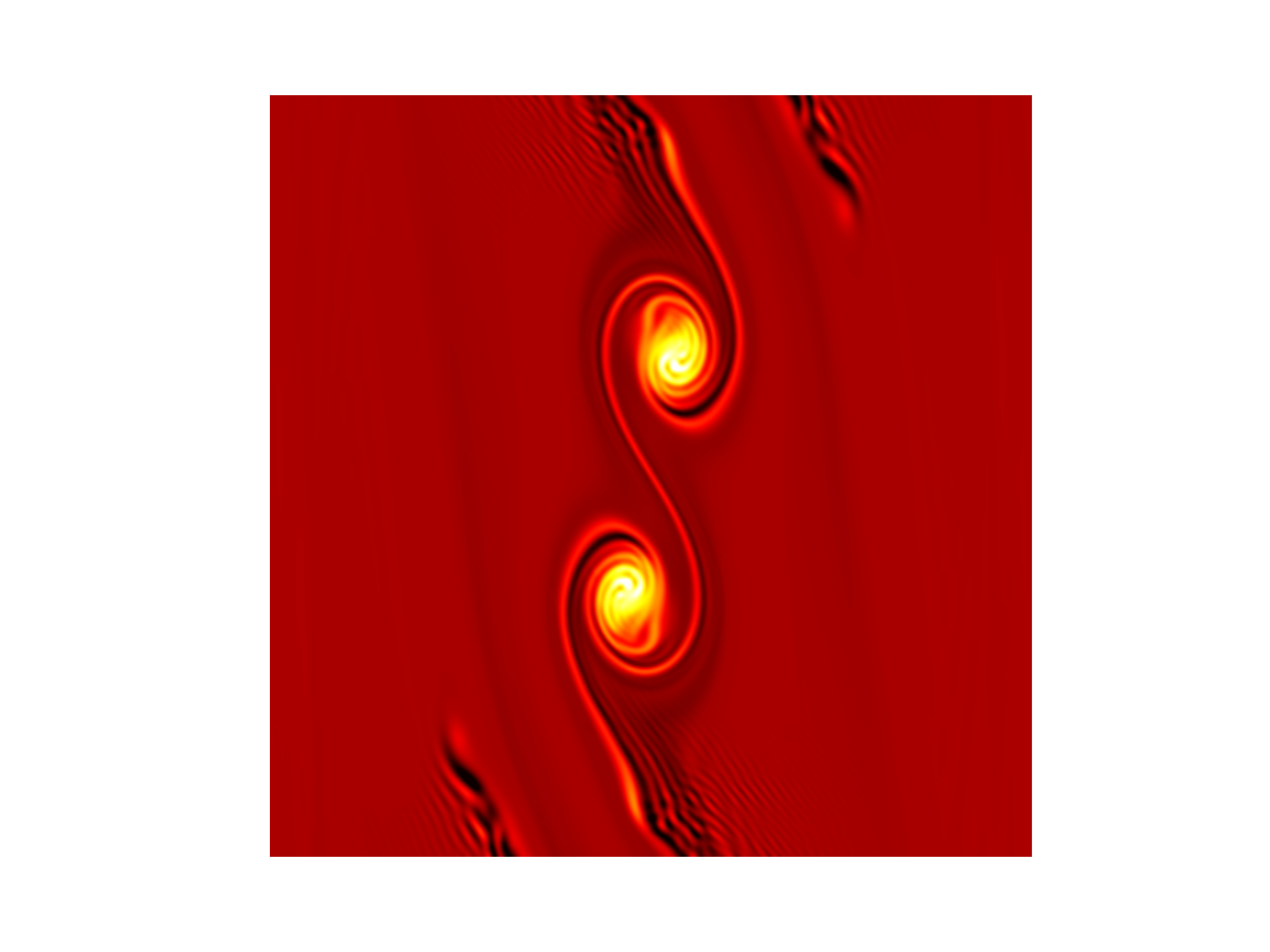}}
\resizebox{0.20\textwidth}{!}{\includegraphics*[trim={4.5cm 1.4cm 4.5cm 2.0cm},clip]{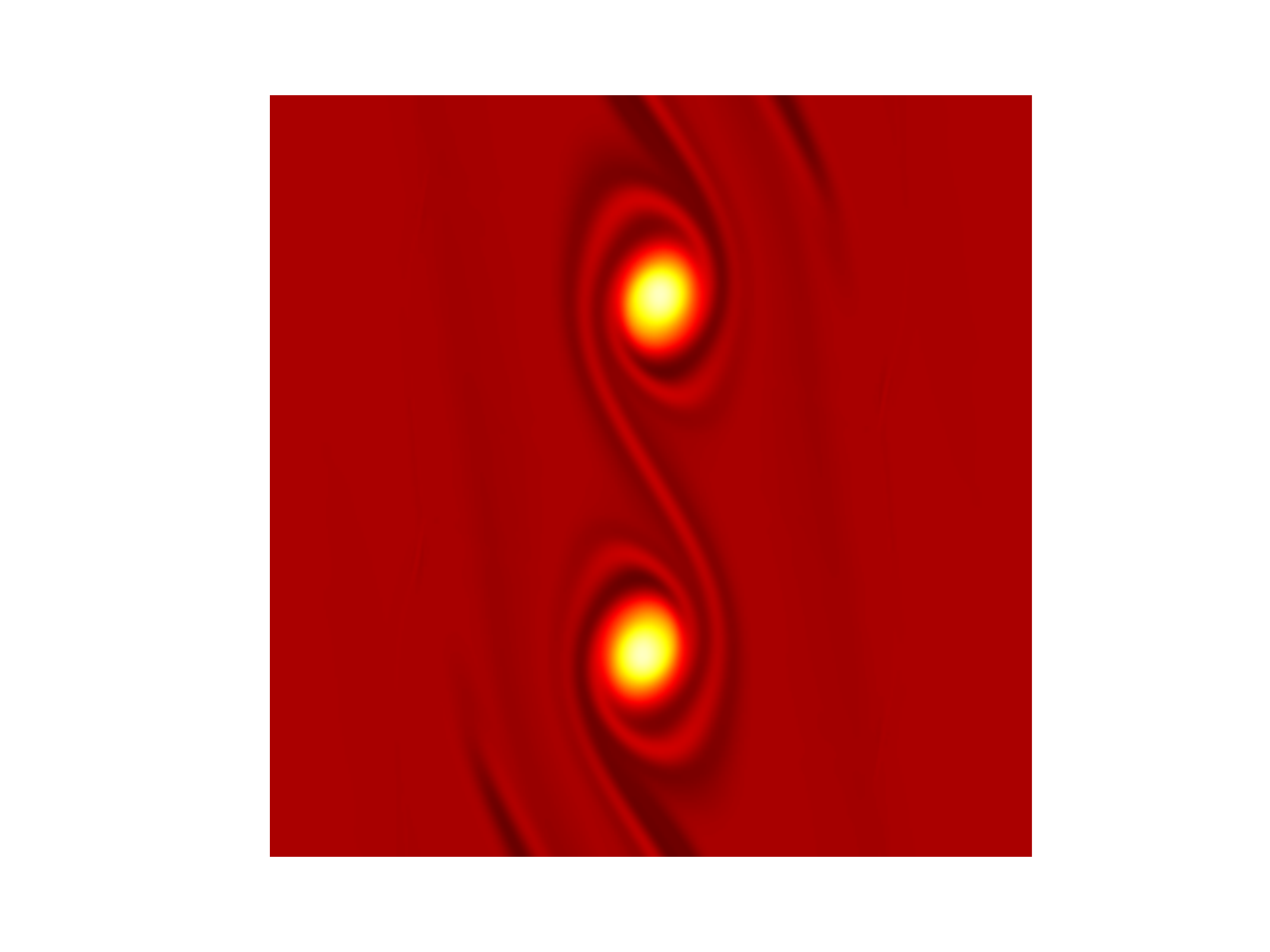}}
\resizebox{0.20\textwidth}{!}{\includegraphics*[trim={4.5cm 1.4cm 4.5cm 2.0cm},clip]{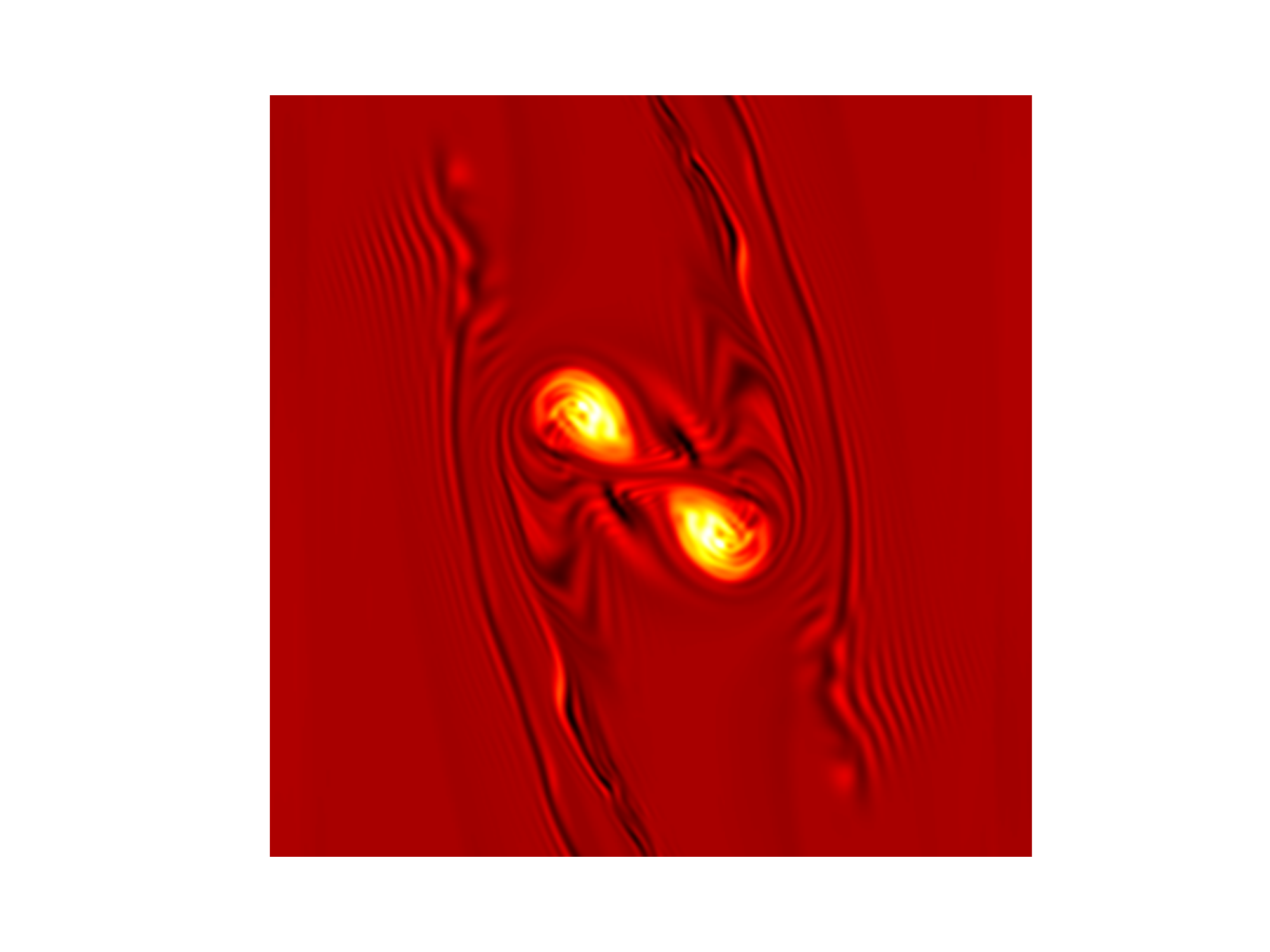}}
\resizebox{0.20\textwidth}{!}{\includegraphics*[trim={4.5cm 1.4cm 4.5cm 2.0cm},clip]{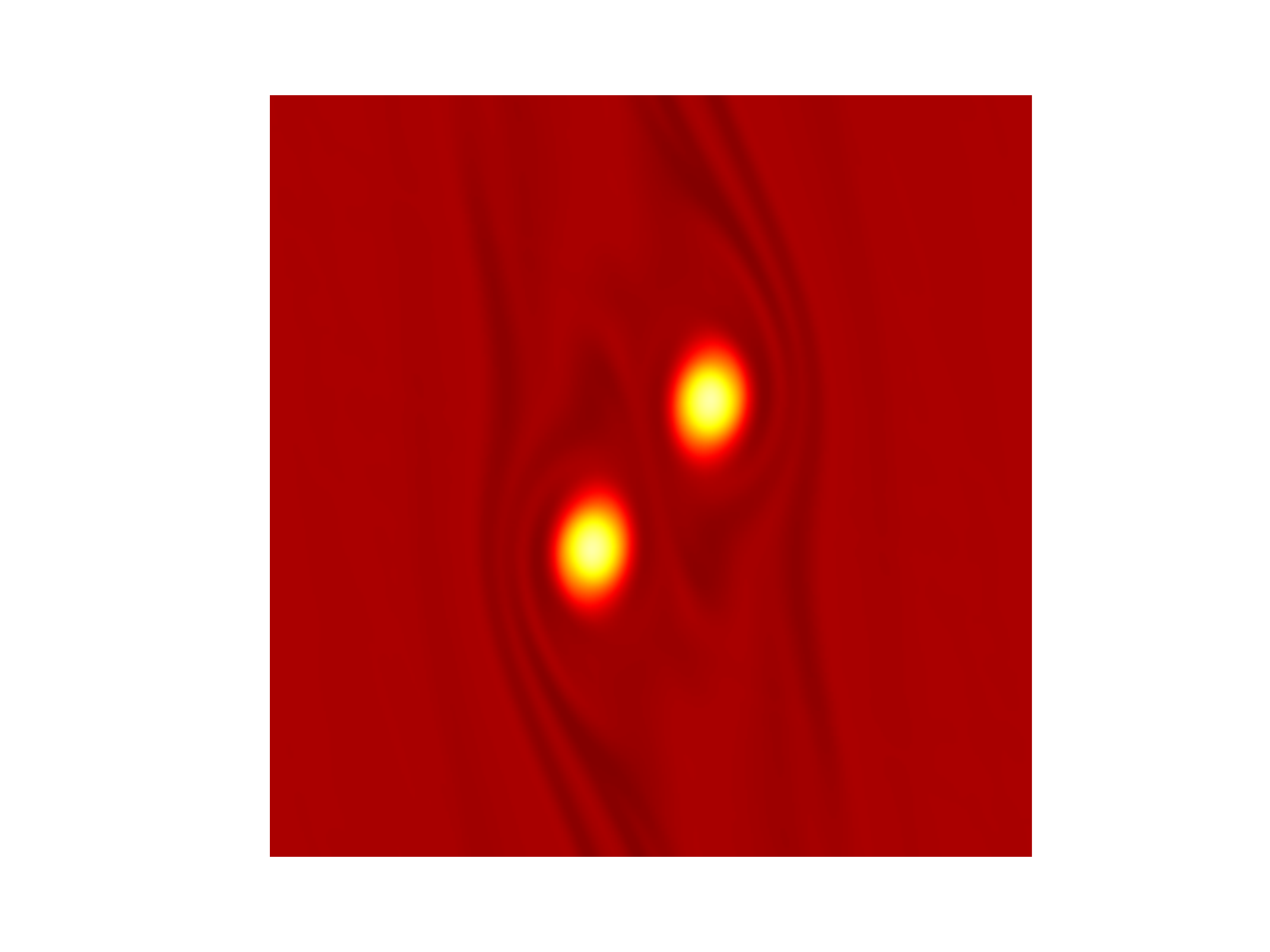}}
\resizebox{0.20\textwidth}{!}{\includegraphics*[trim={4.5cm 1.4cm 4.5cm 2.0cm},clip]{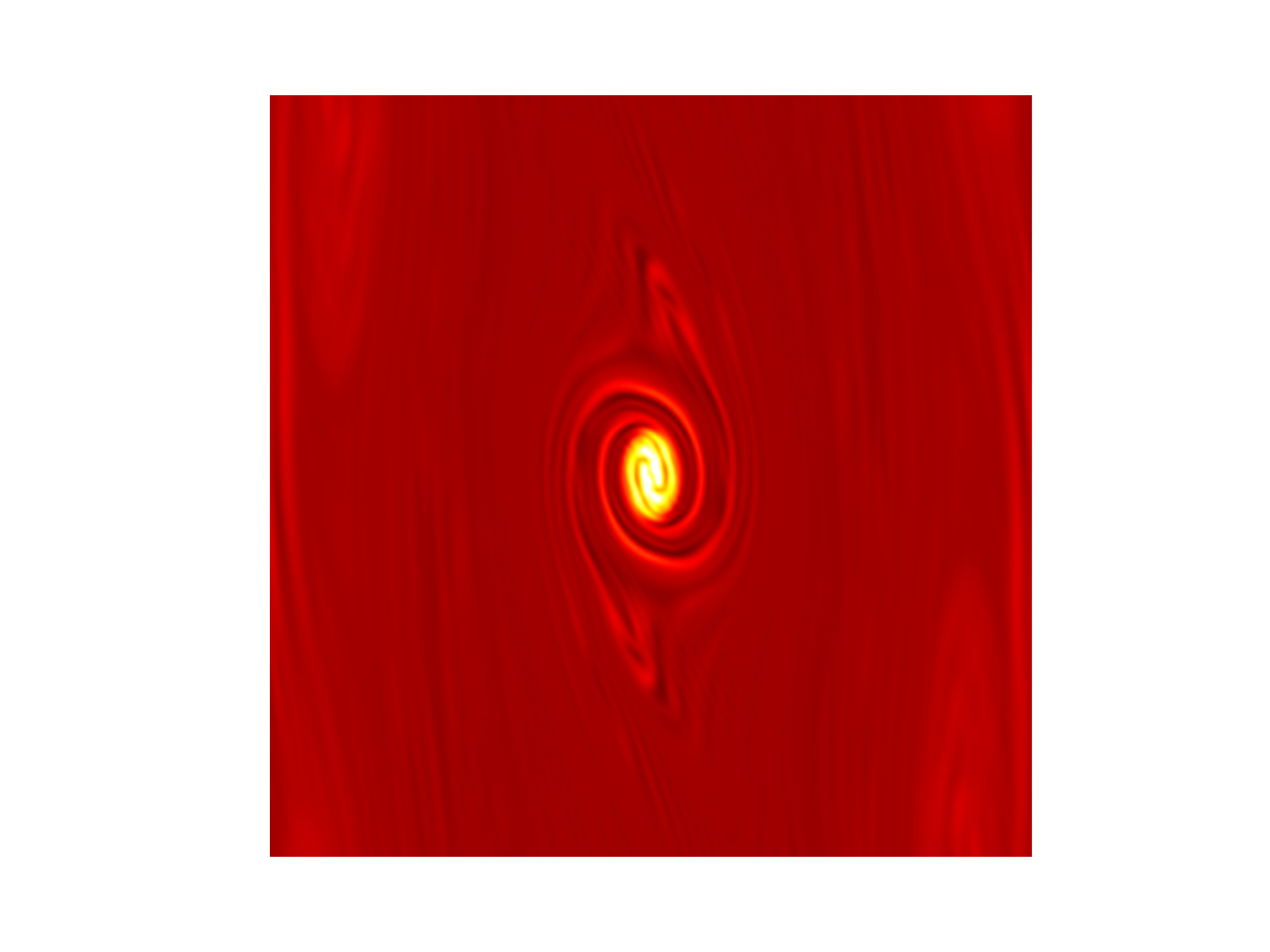}}
\resizebox{0.20\textwidth}{!}{\includegraphics*[trim={4.5cm 1.4cm 4.5cm 2.0cm},clip]{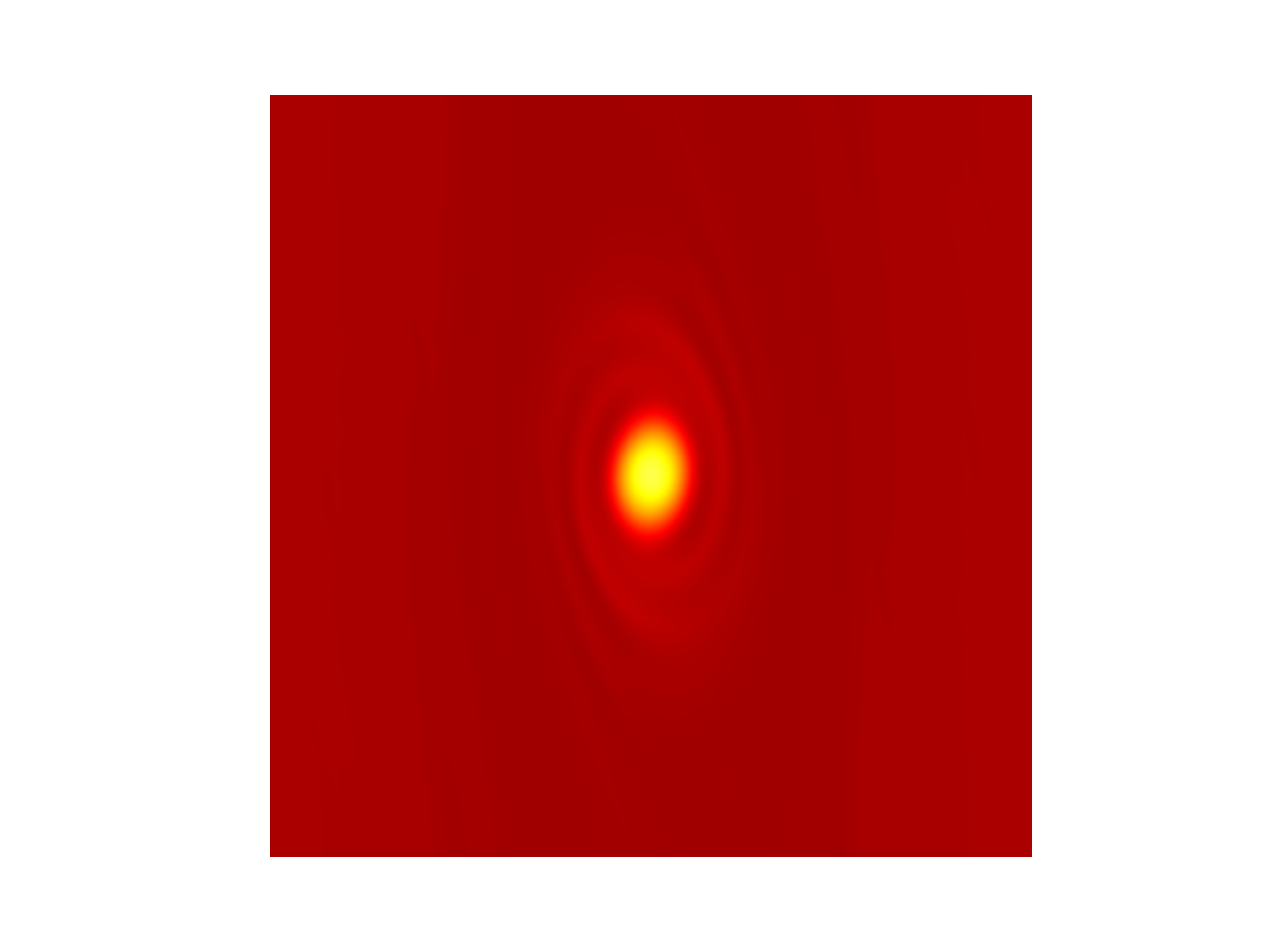}}

\end{center}
\caption{Vorticity in shearing box simulations of two dynamically interacting vortices evolved with RBV2 (left column) and PLUTO (right column).  The simulations were run on 512x512 zones  with an extent of $\Delta R=.04$~AU centered at $R=1$~AU with a sound speed, $c_s=0.05$. The PLUTO snapshots were produced at  times of $0.0$, $2.4$, $4.0$ $6.0$ and $32.8$, from top to bottom, while the RBV2 snapshots were produced at $0.0$, $2.4$, $4.0$ $6.0$ and $55.8$, from top to bottom. Not only is the substructure that develops significantly different, but in RBV2 the vortices complete fourteen orbits before merger, while in PLUTO, they merge after seven orbits.}
\label{fig:dynamicalvortices}
\end{figure}

With these motivations, we seed two identical vortices  into the same Keplerian shearing sheet geometry adopted in the previous subsection. The vortices are centered at $x=\pm0.0025AU$ and $y=\pm0.005AU$ relative to the grid center at the fiducial radius.  As before, the domain contains 512x512 zones with $\Delta R=.04$~AU and  $c_s=.050$.  We run these initial condition in RBV2 and PLUTO  and show snapshots of the vorticity in the simulations in Figure \ref{fig:dynamicalvortices}. The left column shows the dynamical vortex merger in RBV2 at simulation times of $0.0$, $2.4$, $4.0$ $6.0$ and $55.8$, from top to bottom, while the right column shows the same process evolved with PLUTO at simulation times of $0.0$, $2.4$, $4.0$ $6.0$ and $32.8$, from top to bottom. Note the difference in final plotted times for the last pair of frames.  

The dynamical interaction of these vortices in the presence of the Keplerian shear flow as rendered with RBV2 is striking. The vortices initially drift with the shear -- upwards for the left vortex and downwards for the right vortex. However, their mutual attraction overwhelms the shear, and the vortices deform as they rotate towards each other, while both spinning around their own centers and shedding rossby waves that create finely detailed substructure in the flow. The vortices orbit each other fourteen times in the RBV2 simulation, spending most of the simulation time further apart, and rapidly accelerating when they experience near-collisions. The orbital period decreases until the pair finally coalesce into a single uniformly spinning cyclonic disturbance. 

In concordance with the other tests that we have carried out, there appear to be two fundamental differences between the flows evolved with RBV2 and PLUTO. The first is that the substructure that forms in the flow due to the shedding of Rossby waves seems to be more intricate when computed with RBV2. The second, and perhaps more pronounced, discrepancy is that the dynamical merger happens at a substantially later simulation time in RBV2. In the RBV2 calculation, the vortices merge after 14 vortex orbits at a simulation time of $\tau=55.8$, while in PLUTO they merge after 7 orbits at a simulation time of $\tau=32.8$. This delay is consistent with the hypothesis that, for given grid resolution, RBV2 is  running at a higher effective Reynolds number than PLUTO.

\section{Conclusion} 
Nonlaminar flow that exhibits long-lived vortices and time-dependent flow patterns may be an integral feature of both the planet formation and the disk accretion processes, and indeed, flows with these morphologies are ubiquitous in every instance where highly resolved images are available. As a consequence, numerical simulations of disk flows in the strongly nonlinear regime should be designed to respect the conservation of vorticity. In our series of demonstration simulations, we have shown that the explicit van Leer advection methods (and 2nd-order explicit methods generally) suffer from severe, near-immediate vortex dissipation. More unexpectedly, it also appears that higher-order Godunov schemes (such as that implemented in the PLUTO code) also experience significant difficulties in propagating vortical motion. Additionally, as shown by the tests of Lerat et al. 2007, high-order schemes, while far less diffusive than van Leer advection, do not explicitly respect Kelvin's Circulation Theorem as a requirement of their design, and as a consequence show non-negligible numerical degradation over tens of vortex sound-crossing times. Part of this departure may be a product of history. The development of hydrodynamical codes for astrophysical applications had an early heritage largely in the simulation of explosions, and as a consequence, a great deal of attention has been paid to the manner in which the codes handle shocks and advect sharp fronts and discontinuities.

The zeroth-order differences in behavior that we have observed suggest that certain problems, including disk-driven planet migration, and hydrodynamically mediated particle agglomeration might be profitably revisited with simulations in which dissipation is kept to an absolute minimum and vorticity is explicitly conserved.

This paper is not the only one to explore this line of inquiry. Other workers are also pursuing numerical algorithms that exhibit low dissipation, improved vorticity conservation, and finely detailed eddying, e.g. \citet{Springel2010}. The GIZMO code \citep{Hopkins2015}, uses new Lagrangian Godunov-type methods to incorporate advantages of both smoothed-particle hydrodynamical and adaptive mesh refinement schemes, and is able to render test problems such as the Kelvin-Helmholtz instability with impressive resolution and detail. 

In a series of upcoming papers, we plan to use RBV2 to investigate two-dimensional turbulence in accretion disks, disk-driven planet migration, Jupiter's atmosphere, and hurricanes in protostellar disks. We will also explore the code's ability to simulate the Kelvin-Helmholtz and related instabilities, which may manifest to a significant degree in protoplanetary disks \citep{Lee2010}. 

\section{Acknowledgements}
D. S. thanks the Gruber Foundation and Patricia Gruber for the Gruber Fellowship, which supported the work reported here. This material is also based upon work supported by the National Aeronautics and Space Administration through the NASA Astrobiology Institute under Cooperative Agreement Notice NNH13ZDA017C issued through the Science Mission Directorate. We acknowledge support from the NASA Astrobiology Institute through a cooperative agreement between NASA Ames Research Center and Yale University. We thank Fred Adams, Sarbani Basu, Konstantin Batygin, Paolo Coppi, John Forbes, Hubert Klahr, Daisuke Nagai, J.J. Zanazzi, and Robert Zinn  for useful discussions. We also thank the anonymous referee for an excellent report and useful scientific insight. 
\acknowledgements

\end{document}